# Selective profiling of non-canonical nucleic acid structures via size-discriminative supramolecular probes

**Runyu Shi,[1] Dan Huang,[1,2] Yanxi Wang,[1] Qiuju Zhou,[3] Zhenzhen Zhao,[2] Binwu Ying,[2] Qianfan Yang*[1] and Feng Li*[1]**

## Affiliations

**[1]Key Laboratory of Green Chemistry and Technology of Ministry of Education, College of Chemistry Sichuan University, Chengdu, Sichuan, 610064, P. R. China**

**[2]Department of Laboratory Medicine and National Clinical Research Center for Geriatrics, West China Hospital of Sichuan University, Chengdu, Sichuan, 610041, P.R. China**

**[3]College of Chemistry and Chemical Engineering Xinyang Normal University, Xinyang, Henan, 464000, P. R. China**

## Contributions

R. Shi, D. Huang, Y. Wang, and Q. Zhou contributed the acquisition, analysis, and interpretation of data; D. Huang, Z . Zhao and B. Ying collected blood samples from healthy individuals and cancer patients; R. Shi and Q. Yang conceived idea and designed all experiments; R. Shi, Q. Yang and F. Li contributed to writing and revising the manuscript. R. Shi and D. Huang contributed equal to this article. All authors reviewed and approved the final manuscript.

## Corresponding author

Correspondence to:
Associate Professor: Qianfan Yang
Key Laboratory of Green Chemistry and Technology of Ministry of Education, College of Chemistry
Sichuan University
Chengdu, Sichuan, 610064, P. R. China
E-mail: yangqf@scu.edu.cn
ORCID: 0000-0003-4205-1909
Professor: Feng Li
Key Laboratory of Green Chemistry and Technology of Ministry of Education, College of Chemistry
Sichuan University
Chengdu, Sichuan, 610064, P. R. China
E-mail: windtalker_1205@scu.edu.cn
ORCID: 0000-0002-2616-5343

## Abstract

Nucleic acids can form diverse non-canonical structures, such as G-quadruplexes (G4s) and i-motifs (iMs), which are critical in biological processes and disease pathways. This study presents an innovative probe design strategy based on groove size differences, leading to the development of BT-Cy-1, a supramolecular cyanine probe optimized by fine-tuning dimer "thickness". BT-Cy-1 demonstrated high sensitivity in detecting structural transitions and variations in G4s and iMs, even in complex environments with excess dsDNA. Applied to clinical blood samples, it revealed significant differences in RNA G4 and iM levels between liver cancer patients and healthy individuals, marking the first report of altered iM levels in clinical samples. This work highlights a novel approach for precise nucleic acid structural profiling, offering insights into their biological significance and potential in disease diagnostics.

## Introduction

Non-canonical nucleic acid structures, such as G-quadruplexes (G4s)[1] and i-motifs (iMs)[2], are integral to numerous biological processes[3], including DNA and RNA replication[4,5], transcription regulation[6,7], telomere maintenance[8], and RNA splicing.[9] Selective profiling of different non-canonical structures using tools, such as small molecular binders, is critical for understanding their specific roles in cellular function and disease pathogenesis.[10–12] Nevertheless, of more than 4500 G4 ligands and 280 iM ligands documented by far, fewer than 3% exhibit selectivity to non-canonical over canonical structures. Moreover, fewer than 10 ligands are capable of distinguishing iM from G4 structures.[13,14]

To address the challenge of structural selectivity, efforts have been made to engineer supramolecular probes that recognize G4 or iM structures through dynamic interactions. For example, Enoch et al.[15] designed a β-cyclodextrin-functionalized complex that selectively bind specific G4 conformations through size-exclusion and complementary interactions. Hooley and Zhong[16,17] introduced a host-guest sensing system incorporating fluorescent probes and synthetic cavity molecules to distinguish subtle structural variations in G4 topologies. Cyanine dyes, such as ETC[18], MTC[19], SCY-5[20], K21[21], and CV2[22], have also been employed for designing selective supramolecular probes for G4 structures via aggregation manipulation or excimer transitions. Despite the rapid development in supramolecular probes, most designs focused primarily on G4 structures. Probes that can selectively discriminate G4 and iM structures are highly demanded but remain unavailable so far. Herein, we report the design of size-discriminative supramolecular probes capable of selectively profiling G4 and iM structures in buffer and clinical samples.

The idea of size-discriminative supramolecular probes is one the basis of structural analysis of G4 and iM. By analyzing the binding modes of existing ligands that could selectively recognize G4 or iM, we found many were groove binders[23–26], underscoring the critical role of groove structure in ligand binding specificity. We further analyzed groove sizes of varying nucleic acid structures documented in the Protein Data Bank (PDB) and found that G4s had the largest grooves (~16 Å), followed by iMs (~14 Å), and dsDNA (~11.6 Å) (Figure 1). Consistent with previously reported values in the literature[27,28], our structural analysis directed the design of size-discriminative probes.

# Results

## Design and Characterization of BT-Cy Probes

To match the groove size differences between G4s and iMs, we designed a series of benzothiazole Cy3-based probes featuring varying meta-substituents on the cyanoethyl chain (Figure 1, ESI section 2). These probes differ in their meta-substituents: hydrogen (BT-Cy-0), methyl (BT-Cy-1), and ethyl (BT-Cy-2). The dimer formation of each probe was confirmed by NMR analysis (ESI, section 3). The design rationale lies in the steric effects of these substituents, which extended outward from the aromatic rings due to steric hindrance. This structural variation is expected to influence the "thickness" of the dimers, with larger substituents leading to thicker dimers. Consistently, molecular docking simulations[29,30] quantified dimer thicknesses as 9.7 Å for BT-Cy-0, 10.5 Å for BT-Cy-1, and 10.9 Å for BT-Cy-2, respectively, which was further corroborated by NMR DOSY analysis (ESI, section 4). Additionally, binding-mode analysis suggested that the probes act as groove binders, with further details provided in ESI (Section 5).

## Recognition of Different DNA Structures by BT-Cy Probes

We next investigated the monomer-dimer distribution of BT-Cy probes in the presence of G4 (G-VEGF), iM (i-VEGF), or dsDNA (calf thymus DNA) (Table S1). The absorption spectra (Figure 2a-c and S5) revealed two peaks for all three probes, corresponding to the monomer and dimer forms, confirming a monomer-dimer equilibrium in solution31. Upon interaction with different DNA structures, re-establishment of the monomer and dimer equilibrium was observed, as indicated by changes in the monomer-to-dimer absorbance ratio (Abs M/D) (Figure 2d). Notably, G4s induced a pronounced shift toward the dimeric form (reduce in Abs M/D), likely driven by their larger groove sizes compared to iMs and dsDNA. The preference of G4 groove to accommodate the dimeric form of the probes was further supported by the observed fluorescence enhancement in Figure 3a and ESI (section 8).

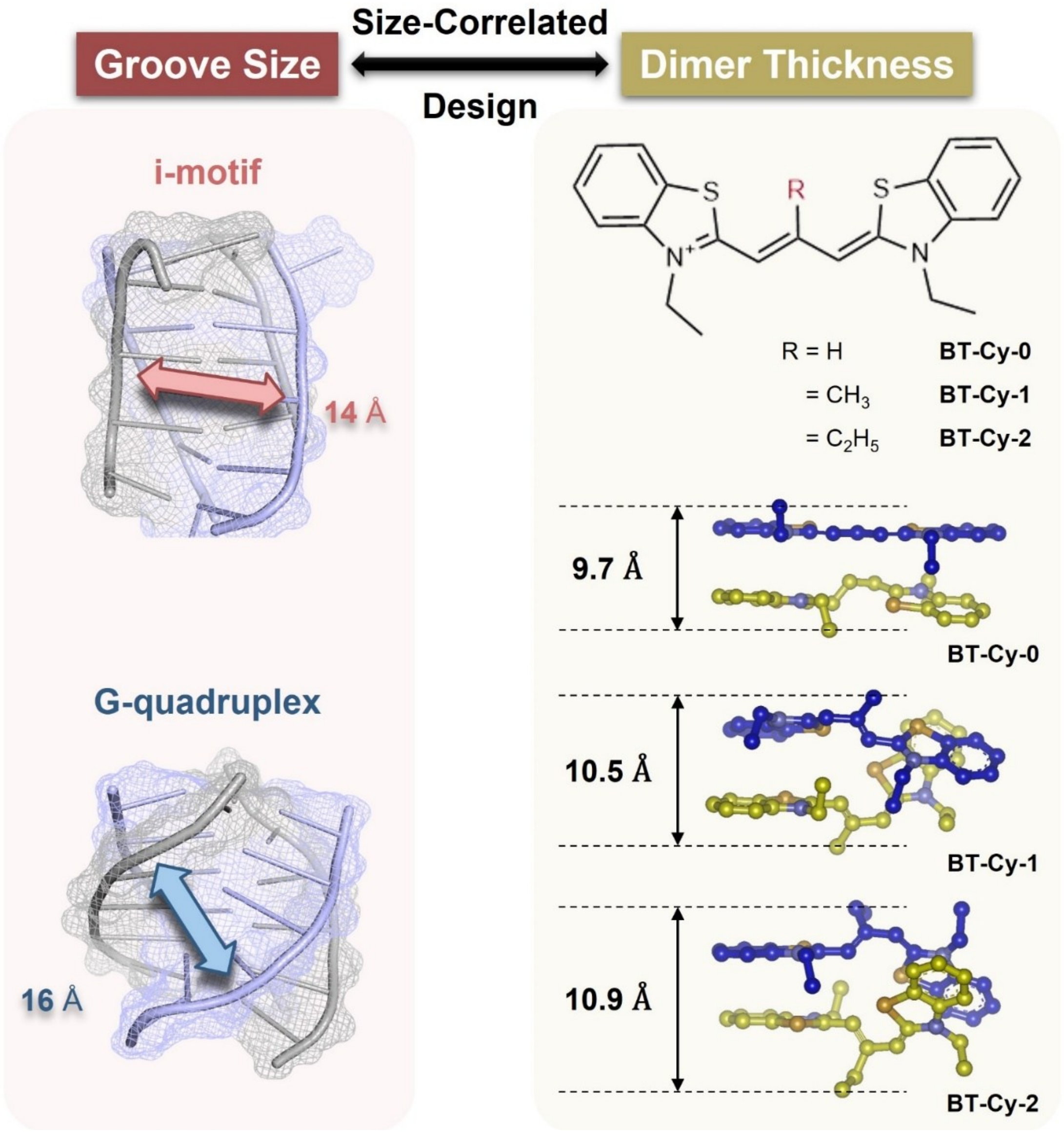

**Fig. 1** Design strategy based on nucleic acid groove size.The groove size differences between iMs (~14 Å) and G4s (~16 Å) guided the design of the BT-Cy series probes with varying dimer "thickness".

To gain a comprehensive understanding of the binding behaviour, we further examined the fluorescence ratio (M/D ratio) between monomeric and dimeric forms of probes against various DNA structures, including G4s, iMs, dsDNA, and ssDNA (Figure 3b and ESI section 8). The analysis revealed that BT-Cy-1 and BT-Cy-2 exhibited distinct M/D ratios across different classes of DNA structures with iMs induced highest M/D ratio, suggesting that the monomeric form of the probes bound preferentially to iMs. By contrast, BT-Cy-0 showed no structural specificity.

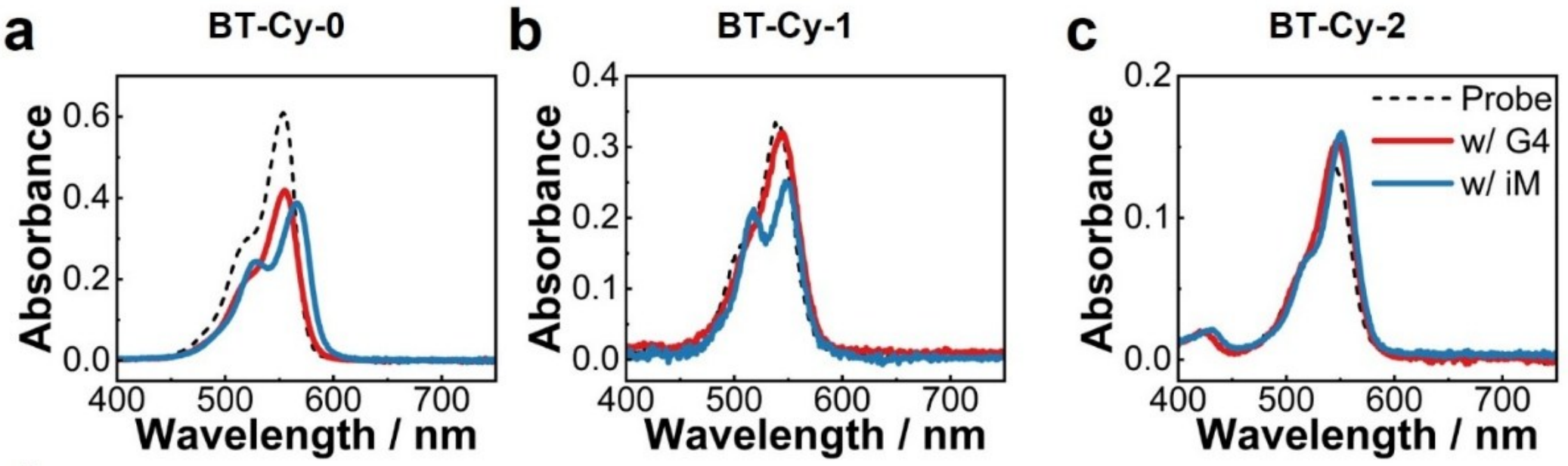

d

| Abs M/D | Probe | w/ G4 | w/ iM | w/ dsDNA |
|---|---|---|---|---|
| **BT-Cy-0** | 2.005 | 1.591 (-20.68 %) | 2.031 (+1.25 %) | 1.242 (-38.04 %) |
| **BT-Cy-1** | 2.072 | 1.184 (-42.84 %) | 1.765 (-14.78 %) | 1.768 (-14.65 %) |
| **BT-Cy-2** | 2.240 | 2.241 (+0.02 %) | 2.306 (+2.93 %) | 2.448 (+9.27 %) |

**Fig. 2** Influence of different DNA structures on the monomer-dimer distribution of BT-Cy probes. Absorption spectra of the probes (5 μM) in the presence of G4 and iM (5 μM) are shown for **a** BT-Cy-0, **b** BT-Cy-1, and **c** BT-Cy-2. **d** The monomer-to-dimer absorbance ratio (Abs M/D) and percentage changes for each probe upon binding to different DNA structures.

Building on the observation that dimers bound preferentially to G4s with enhanced fluorescence and monomers bound preferentially to iMs with higher M/D ratios, we constructed a two-dimensional (2D) plot with y-axis reflecting dimer fluorescence and x-axis reflecting M/D ratios, respectively. As shown in Figure 3c, BT-Cy-0 demonstrated poor selectivity to varying nucleic acid structures, while both BT-Cy-1 and BT-Cy-2 could distinguish among G4, iM, and ds/ssDNA, evidenced by the distribution of different classes of structures on the 2D plot. To quantitatively assess the probes' ability to differentiate DNA structures, we performed support vector machine (SVM) analysis to determine thresholds delineating different zones of nucleic acid structures and then calculated classification separation scores (C-scores). BT-Cy-1 achieved the highest C-score, followed by BT-Cy-2 and BT-Cy-0. Collectively, our results revealed that BT-Cy-1 was the best probe demonstrating superior ability to selectively profile different classes of nucleic acid structures.

The observed poor selectivity of BT-Cy-0 was consistent with our hypothesis that BT-Cy-0 formed thinnest dimer and thus could not sense groove size differences among G4s, iMs, and dsDNA. On the other hand, by forming the thickest dimer, BT-Cy-2 showed better differentiation in the M/D ratio but exhibited weaker overall binding affinity to DNA structures, resulting in minimal dimer fluorescence changes. Together, these factors led to its limited separation ability on the 2D plot. BT-Cy-1, with a proper dimer thickness, achieved the optimal balance between DNA binding affinity and monomer-dimer distribution sensitivity, leading to desired selectivity to nucleic acid structures.

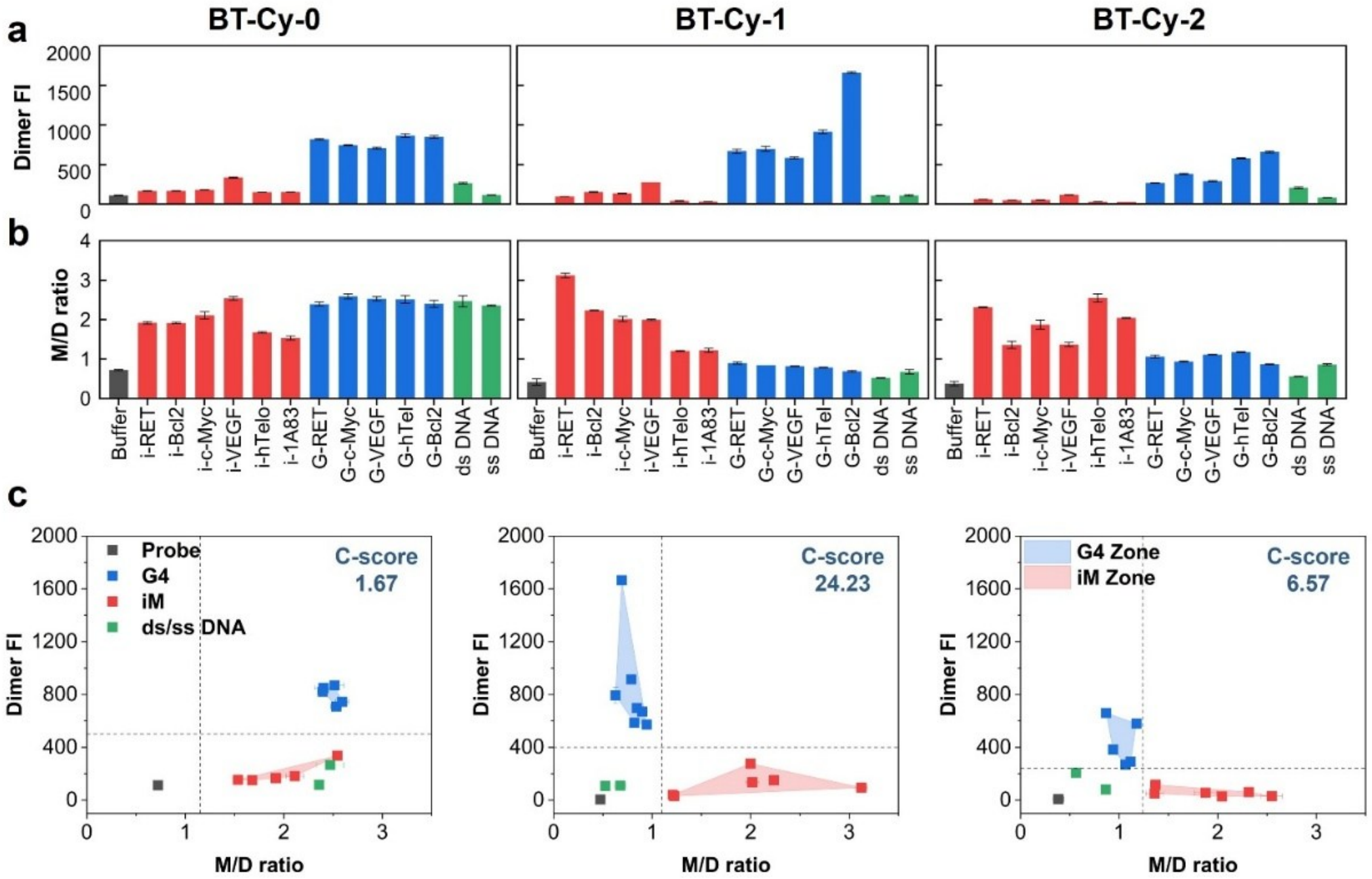

**Fig. 3** Evaluation of BT-Cy probes' ability to differentiate between DNA structures using fluorescence channels. **a** Dimer fluorescence intensity (Dimer FI) and **b** monomer-to-dimer fluorescence ratio (M/D ratio) of BT-Cy-0, BT-Cy-1, and BT-Cy-2 in the presence of various DNA structures. **c** 2D plots of Dimer FI versus M/D ratio for BT-Cy probes, illustrating the separation of different DNA structures into distinct zones. The thresholds used to delineate the zones (dashed lines) and the comprehensive classification separation scores (C-scores) were both derived from SVM calculations.

To further challenge BT-Cy-1 to profile subtle changes in nucleic acid structures, we examined its response to structural transitions induced by environmental factors and sequence modifications. Using the human telomeric G4 sequence as a model, we gradually increased [K+] to induce a transition from ssDNA to a stable hybrid G4 structure (ESI, section 9). The results in Figure 4a revealed a linear shift of the 2D signals from the ssDNA to the G4 one. We also modified a VEGF iM sequence by systematically elongating the length of the groove or loop motifs (ESI, section 9), which were all sensitively profiled on the 2D plot (Figure 4b). Collectively, our results confirmed that BT-Cy-1 is highly sensitive to dynmaic and subtle changes in specific classes of nucleic acid structures.

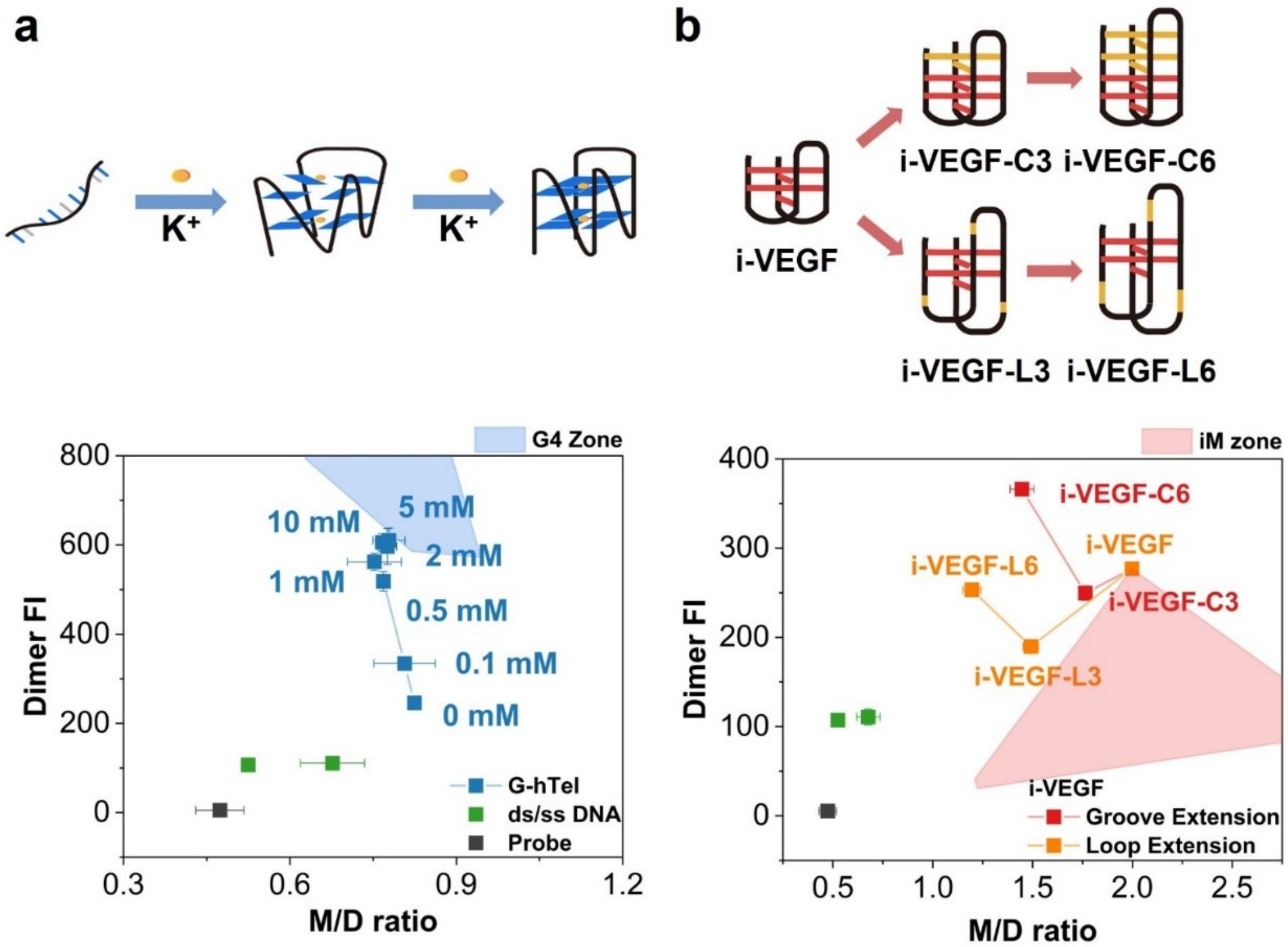

**Fig. 4** Evaluation of BT-Cy-1's capability for nucleic acid structural profiling. **a** A schematic representation of the structural transition of a human telomeric G4 sequence from a single strand to a stable hybrid G4 structure upon increasing [K+] and the corresponding 2D plot. **b** A schematic representation of the structural variations of a VEGF iM sequence induced by systematic elongation of groove and loop length, respectively, and the related 2D plot.

## Clinical Application of BT-Cy-1

We finally challenged BT-Cy-1 to selectively profile nucleic acid structures in clinical samples. Recently, Sun et al. [20] designed a cyanine supramolecular probe SCY-5, which revealed distinct RNA G4 distribution patterns in clinical blood samples, indicating that G4 distributions hold the potential as a diagnostic marker for cancer diagnosis. Here, with our BT-Cy-1, we first confirmed that M/D ratios and dimer fluorescence could be used to quantify the levels of iM and G4 structures, respectively, in the prescence of large access of ds/ss nucleic acids. The lowest detectable of iM and G4 were determined to be 4.75% and 1%, respectively (Figure 5**a** and 5**b**).

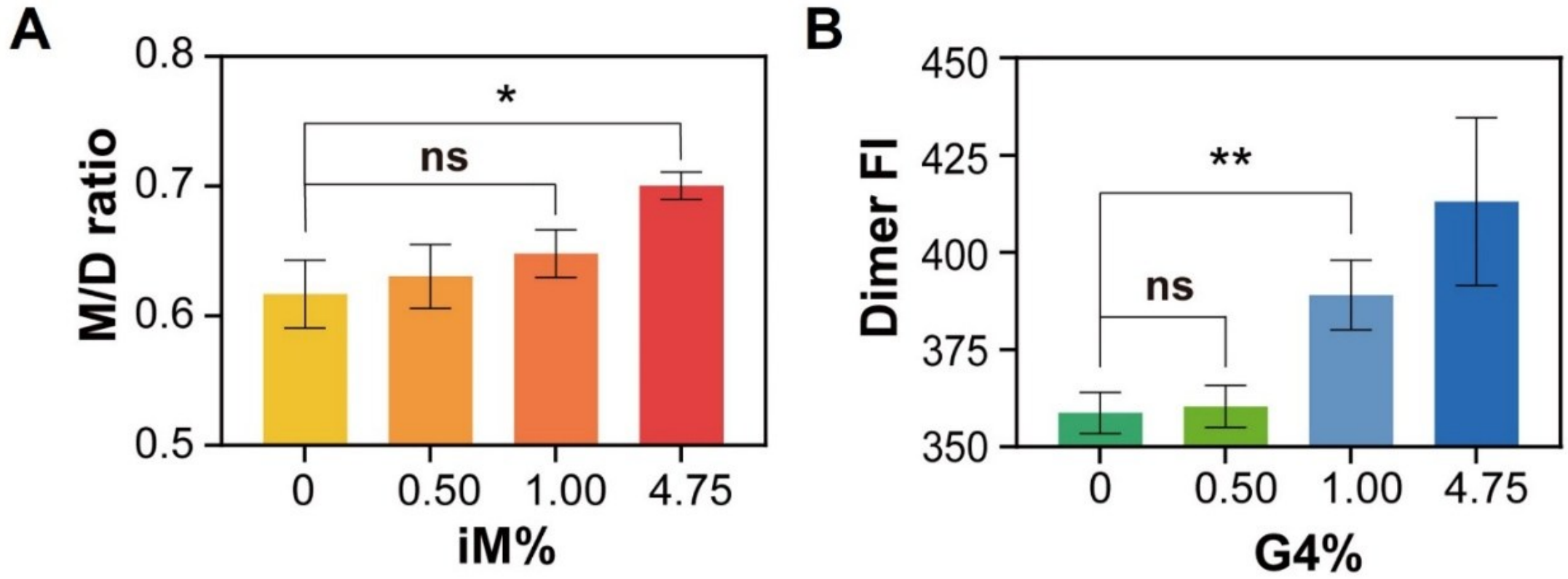

**Fig. 5** Performance of BT-Cy-1 in quantitative detection. Detection limits of **a** iM and **b** G4 in the presence of a large excess of dsDNA background.

We then collected blood samples from liver cancer patients and healthy individuals, extracted total RNA, and analyzed the sample using BT-Cy-1 (Figure 6a). We found that RNA from cancer patients exhibited measurable higher iM levels (n = 15) than healthy individuals (n = 8). Nevertheless, a further analysis using receiver operating characteristic (ROC) curve and confusion matrix revealed clinical sensitivity, specificity, and area under the curve (AUC) to be only 60.0%, 62.5%, and 0.66, respectively (Figure 6b). By contrast, RNA G4 levels of cancer patients (n = 5) were significantly lower than those of healthy individuals (n = 5). More excitingly, when using G4 level as a marker, we achieved 100% for both clinical sensitivity and specificity (Figure 6c). Although the existence of iMs and G4s in biological systems has been well established, investigation of the diagnostic potential of such non-canonical structures remain rare. Using BT-Cy-1 as a structure selective tool, we, for the first time, observed significant changes in both G4 and iM levels in cancer patients, paving a way for future clinical investigations.

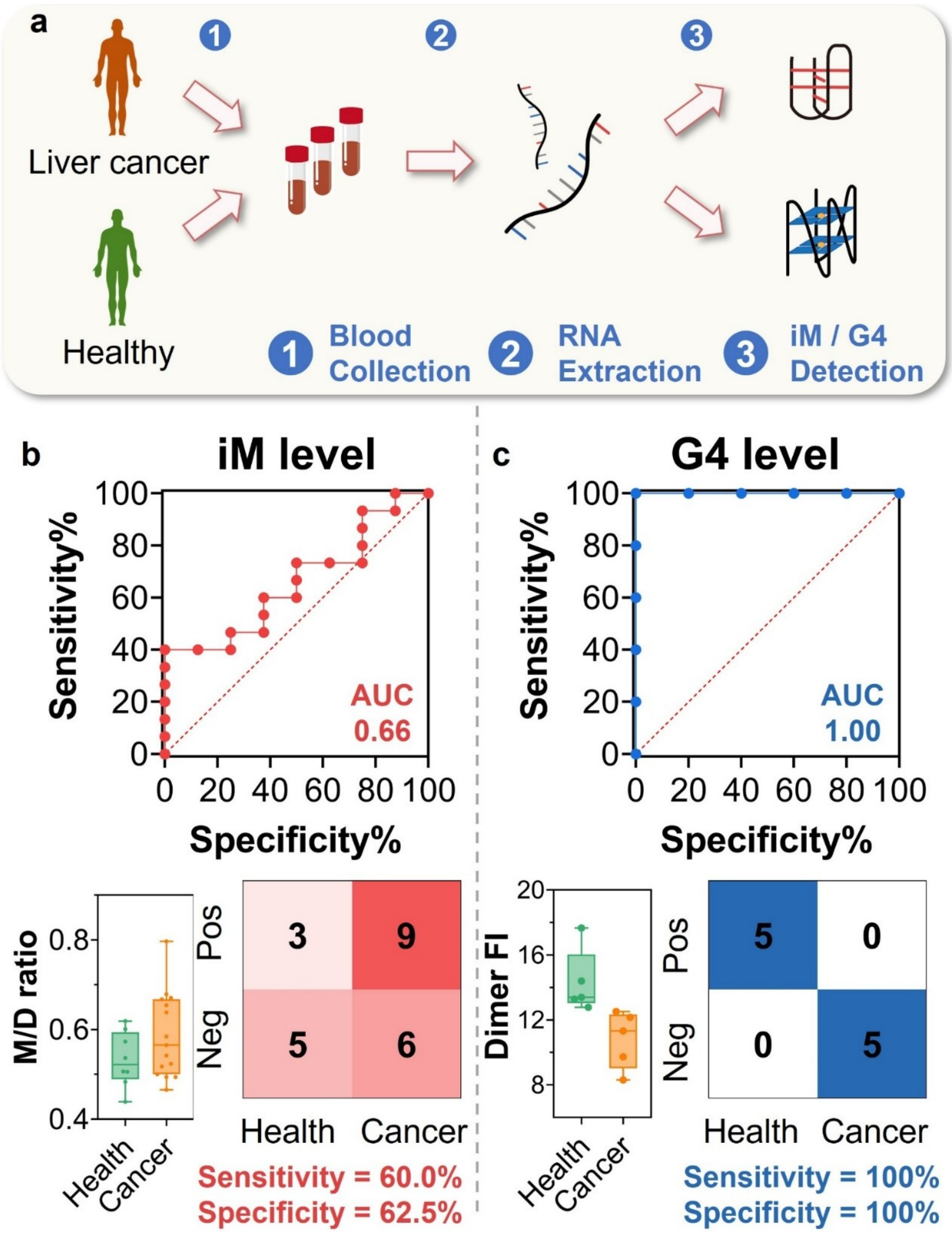

**Fig. 6:** Performance of BT-Cy-1 in clinical applications. **a** Detection workflow of RNA extracted from blood samples of liver cancer patients and healthy individuals using BT-Cy-1. The ROC curves, scatter plots, and confusion matrices for **b** iM and **c** G4 level detection in RNA extracted from blood samples of liver cancer patients and healthy individuals.

## Discussion

In this study, we developed a series of supramolecular cyanine probes based on groove size differences, specifically designed to differentiate between non-canonical nucleic acid structures such as G4s and iMs. By fine-tuning the dimer thickness through structural modifications, BT-Cy-1 achieved an optimal balance between DNA binding affinity and monomer-dimer distribution sensitivity, enabling precise profiling of various nucleic acid structures. Finally, we applied BT-Cy-1 to clinical blood samples and successfully detected significant differences in RNA G4 and iM levels between liver cancer patients and healthy individuals. The probe provided the first-ever evidence of altered iM levels in clinical samples while also confirming the abnormality of G4 levels in cancer patients, offering new insights into the potential roles of both iM and G4 in cancer biology.

# Methods

## Chemicals and materials

All reagents were used as received without further purification. 2-Methylbenzothiazole (Product No. 17230I), triethyl orthoacetate (Product No. 83364N), and triethyl orthopropionate (Product No. 15693F) were purchased from Adamas. Iodoethane (Product No. 80977A) and triethyl orthoformate (Product No. 17934A) were purchased from Sigma-Aldrich. All DNA oligonucleotides were obtained from Sangon Biotech Co., Ltd. (Shanghai, China) and purified using HAP (details in Table S1). Calf thymus (CT) DNAs (double-stranded: Product No. D4522; single-stranded: Product No. D8899) were purchased from Sigma-Aldrich.

## Molecular Structure Identification

$^{1}$H-NMR spectra and NOESY were measured on an Avance II-400 MHz spectrometer and an Avance II-600 MHz spectrometer in appropriated 500 μL deuterated solution (CD3OD and DMSO-d6). The mess of the sample is 2-10 mg. A high-resolution Q Exactive Plus hybrid quadrupole-Orbitrap mass spectrometer (Thermo Fisher Scientific, San Jose, CA) equipped with a heated electrospray ionization (H-ESI) source was used in this study. The samples were dissolved in LS-MS grade methanol and diluted to 1 μg/mL for injection.

## Dimer Thickness Measurement and Comparation

The dimer thickness of BT-Cy probes was first compared quantitatively using DOSY NMR. NMR measurements were performed on a 600 MHz NMR spectrometer (JEOL ECZ600R/S3) equipped with a 14.09 T superconducting magnet and a 5.0 mm broadband Z-gradient high resolution ROYAL probe (JEOL RESONANCE Inc., Japan). The samples were dissolved in 500 μL CD3OD (containing 0.03% TMS) and then transferred into 5 mm NMR tubes. The 1H DOSY spectra were obtained by using bpp_led_dosy_pfg sequence with the following parameters: x_Freq=600.17 MHz, x_sweep=13 ppm, x_offset=4.2 ppm, x_points=32 K, relaxation_delay=7.0 s, diffusion_time=100 ms, delta=2 ms, and scans=32. The 1H DOSY experiments were conducted at 293 K.

Secondly, molecular docking simulations were performed using Autodock 4.2 software32 to generate the dimer configuration. In this process, one BT-Cy molecule was used as the ligand, and the other as the receptor. The docking simulation was run to predict the optimal binding pose and construct the dimer. The resulting PDB file from the docking calculation was then used for further analysis. Then, the PDB file of the dimer was opened in Multiwfn software. After sequentially entering "100", "21", and "size", the program, under the main function "100", accessed the sub-function "output various structural information". This provided the dimensions of the dimer, including the length, width, and height, based on a Cartesian coordinate system centered on the molecule.

## DNA Secondary Structure and Structural Transition Identification

The C-rich oligonucleotide was added to BR buffer (40 mM acetic acid, 40 mM phosphoric acid, and 40 mM boric acid, pH = 4.5) to achieve a final DNA concentration of 5 μM. The G-rich oligonucleotide was added to PB buffer to obtain a final DNA concentration of 5 μM. For the G-hTel sequence, the oligonucleotide was added to 10 mM Tris-HCl buffer with varying concentrations of KCl (0-10 mM) to achieve a final DNA concentration of 5 μM.

Circular dichroism (CD) spectra were recorded on a Chirascan CS30100 circular dichroism spectrometer (Applied Photophysics, Leatherhead, UK) using a 1 cm quartz cuvette. The scanning range was from 220 to 320 nm, the scanning speed was 1200 nm/min, and the number of scans was set to 3. Baseline correction was performed with the corresponding buffer solution before data collection.

## Monomer-Dimer Distribution of BT-Cy Probes

The Monomer-Dimer distribution of BT-Cy probes was studied by measuring UV-vis absorption spectra using a Thermo Scientific EVOLUTION 201 UV-Visible Spectrophotometer with a 1 cm quartz cuvette.

For sample preparation, i-VEGF was added to BR buffer to achieve a final DNA concentration of 5 μM, with BT-Cy probes at the same concentration. For G-VEGF or dsDNA, these were added to PB buffer to achieve final concentrations of 5 μM for G-VEGF or 36 μg/mL for dsDNA, with BT-Cy probes at the same concentration.

## Differentiation of DNA Structure and Structural Transition

For the C-rich oligonucleotide, it was added to BR buffer to achieve a final DNA concentration of 5 μM, while adding BT-Cy probes at the same concentration. The G-rich oligonucleotide or ds/ssDNA was added to PB buffer to achieve a final DNA concentration of 5 μM for G-VEGF or 36 μg/mL for ds/ssDNA, with BT-Cy probes at the same concentration. For the G-hTel sequence, the oligonucleotide was added to 10 mM Tris-HCl buffer with varying concentrations of KCl (0-10 mM) to achieve a final DNA concentration of 5 μM, and BT-Cy-1 probes were added at the same concentration.

Fluorescence measurements were performed using a fluorescence spectrometer with excitation wavelengths set to 534 nm for dimer fluorescence and 568 nm for monomer fluorescence. The input and output slits were both set to 2.5 nm, and the voltage was adjusted to 700 V.

## Support Vector Machine (SVM) Analysis for Threshold and C-score Calculation

To assess the ability of the BT-Cy probes to differentiate between DNA structures, we performed SVM analysis to determine thresholds separating different DNA structure zones and calculate a comprehensive classification separation score (C-score). The analysis steps are as follows:

Experimental data, including the dimer fluorescence intensity (Dimer FI) and monomer/dimer (M/D) ratio, were used as input features for the SVM model. Each data point was labeled according to its corresponding DNA structure type (e.g., G4, iM, or dsDNA). The dataset was normalized to ensure all features had zero mean and unit variance.

A linear SVM classifier was used to identify decision boundaries that separated the different DNA structure classes. The SVM model solves the following optimization problem:

$$\min_{w,b} \frac{1}{2}\|w\|^2 \; subject\ to\ y_i(w^\top x_i + b) \geq 1, \forall i$$

Where w is the weight vector defining the hyperplane, b is the bias term, x_i represents the feature vector of the ith data point, y_i ∈ {-1, 1} is the calss label of the ith data point.

The resulting hyperplanes defined the thresholds Tij between each pair of DNA structure classes in the feature space.

The thresholds were calculated as the midpoints of the SVM hyperplanes between adjacent classes. For a given pair of classes i and j, the threshold was determined as:

$$T_{ij} = -\frac{b}{||w||}$$

Where b is the bias term, and ||w|| is the norm of the weight vector. These thresholds were used to delineate distinct DNA structure zones.

To evaluate the overall classification performance of the SVM model, we calculated a comprehensive classification separation score (C-score), defined as:

$$C = \frac{1}{K}\sum_{K=1}^{K}\frac{M_K}{\sigma_K}$$

Where K is the number of DNA structure class, MK is the margin width for class k, calculated as the distance between the hyperplane and the nearst support vector, σK is the standard deviation of feature distribution for class k.

The C-score provides a single metric summarizing the separation quality across all DNA structure classes, with higher scores indicating better overall classification performance.

## Quantitative Detection of iM and G4 in the Presence of Excess dsDNA

i-VEGF was added to BR buffer to achieve a final DNA concentration of 0-0.25 μM, with a fixed concentration of dsDNA at 36 μg/mL (calculated at 24 nt, equivalent to 5 μM). BT-Cy-1 was added to a final concentration of 5 μM. For the G-VEGF sample, it was added to PB buffer to achieve a final DNA concentration of 0-0.25 μM, with a fixed concentration of dsDNA at 36 μg/mL, and BT-Cy-1 was added to a final concentration of 5 μM.

Fluorescence measurements were performed using a fluorescence spectrometer, with excitation wavelengths set to 534 nm for dimer flurescence and 568 nm for monomer fluorescence. The input and output slits were both set to 2.5 nm, and the voltage was adjusted to 700 V.

## Extraction and Detection of RNA from Clinical Blood Samples

The clinical study was approved by the Ethics Committee of West China Hospital, Sichuan University (approval number: 2024.169). Total RNA was extracted from 1.5 mL of EDTA-anticoagulated whole blood using the RNAprep Pure Hi-Blood Kit (DP443, Tiangen Biotech (Beijing) Co., Ltd.) following the manufacturer's instructions. The RNA concentration of each sample was measured using a BioTek Cytation5 instrument.

To standardize the RNA concentration, each sample was adjusted to 9 ng/μL by adding either BR buffer (for iM detection) or PB buffer (for G4 detection) in DEPC-treated water. BT-Cy-1 (5 μM) was then added to each sample, and the mixture was incubated for 5 hours. Fluorescence measurements were performed using a fluorescence spectrometer, with excitation wavelengths set to 534 nm and 568 nm. The input and output slits were both set to 2.5 nm, and the voltage was adjusted to 700 V.

## Statistical information

All data are presented as mean ± standard deviation (SD) unless otherwise noted. The significance of differences between groups was determined using Student's t-test for pairwise comparisons. A p-value of $< 0.05$ was considered statistically significant. Blood samples from 20 healthy individuals and 13 liver cancer patients were collected for RNA extraction. Among them, RNA from 15 healthy individuals and 8 cancer patients was used for iM detection, and RNA from 5 healthy individuals and 5 cancer patients was used for G4 detection

## Data Availability

All other relevant data are available upon reasonable request. Source data are provided with this paper.

## Acknowledgements

This work was supported by National Natural Science Foundation of China [22077087, 22474082], Sichuan Science and Technology Program [2025NSFJQ0019] and The Nanhu Scholars Program for Young Scholars of XYNU.

## Ethics declarations

All human participants provided written informed consent before participating in the study. The study was approved by the Ethics Committee of West China Hospital of Sichuan University, No.2024.169, and it was conducted in accordance with the Declaration of Helsinki.

## Competing interests

The authors declare no competing interests.

# Selective profiling of non-canonical nucleic acid structures via size-discriminative supramolecular probes

Runyu Shi,[1] Dan Huang,[1,2] Yanxi Wang,[1] Qiuju Zhou,[3] Zhenzhen Zhao,[2] Binwu Ying,[2] Qianfan Yang*[1] and Feng Li*[1]

[1] Key Laboratory of Green Chemistry and Technology of Ministry of Education, College of Chemistry Sichuan University, Chengdu, Sichuan, 610064, P. R. China

[2] Department of Laboratory Medicine and National Clinical Research Center for Geriatrics, West China Hospital of Sichuan University, Chengdu, Sichuan, 610041, P.R. China

[3] College of Chemistry and Chemical Engineering Xinyang Normal University, Xinyang, Henan, 464000, P. R. China

## Table of contents

## 1. DNA oligonucleotides used in this work

**Table S1** DNA oligonucleotides used in this work.

| Sequence name | Sequence (5' to 3') | Bases (nt) | Topology |
|---|---|---|---|
| **i-VEGF** | GACCCCGCCCCCGGCCCGCCCCGG | 24 | iM |
| **i-RET** | CCCCGCCCCGCCCCGCCCCTA | 21 | iM |
| **i-hTel** | CCCTAACCCTAACCCTAACCCT | 22 | iM |
| **i-1A83** | CCTTTCCTTTACCTTTCC | 18 | iM |
| **i-c-Myc** | TTACCCACCCTACCCACCCTCA | 22 | iM |
| **i-Bcl2** | CCCGCCCAATTCCTCCCGCGCCC | 23 | iM |
| **i-hTel-1-Br** | $^{Br}$CCCTAACCCTAACCCTAACCCT | 22 | iM |
| **i-hTel-2-Br** | C$^{Br}$CCTAACCCTAACCCTAACCCT | 22 | iM |
| **i-hTel-3-Br** | CC$^{Br}$CTAACCCTAACCCTAACCCT | 22 | iM |
| **i-hTel-7-Br** | CCCTAA$^{Br}$CCCTAACCCTAACCCT | 22 | iM |
| **i-hTel-8-Br** | CCCTAAC$^{Br}$CCTAACCCTAACCCT | 22 | iM |
| **i-hTel-9-Br** | CCCTAACC$^{Br}$CTAACCCTAACCCT | 22 | iM |
| **i-VEGF-C3** | GACCCCCCCGCCCCCCCCGGCCCCCCGCCCCCCCGG | 36 | iM |
| **i-VEGF-C6** | GACCCCCCCCCCGCCCCCCCCCCCGGCCCCCCCCCGCCCCCCCCCCGG | 48 | iM |
| **i-VEGF-L3** | GACCCCGCTGACCCTGACGGCCCGCTGACCCGG | 33 | iM |
| **i-VEGF-L6** | GACCCCGCTGATGACCCTGATGACGGCCCGCTGATGACCCGG | 42 | iM |
| **G-VEGF** | CCGGGGCGGGCCGGGGGCGGGGTC | 24 | Parallel G4 |
| **G-RET** | TAGGGGCGGGGCGGGGCGGGG | 21 | Parallel G4 |
| **G-c-MYC** | TGAGGGTGGGTAGGGTGGGTAA | 22 | Parallel G4 |
| **G-Bcl2** | GGGCGCGGGAGGAATTGGGCGGG | 23 | Hybrid G4 |
| **G-hTel** | TTAGGGTTAGGGTTAGGGTTAGGG | 24 | Hybrid G4 |

## 2. Synthesis and Characterization of BT-Cy Probes

The probes were synthesized according to the methods of Hamer[1] and Ficken[2], including two steps, shown at Scheme S1 and Scheme S2, respectively.

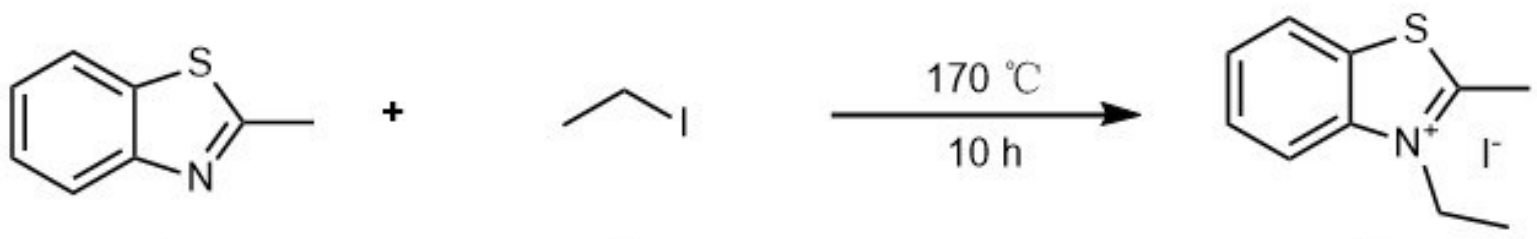

1 2 3

**Scheme S1** Synthetic scheme of compound 3

**3-ethyl-2-methylbenzo[d]thiazol-3-ium iodide (Compound 3)**

A mixture of compound 1 (1.6424 g, 11.01 mmol) and ethyl iodide (2.7558 g, 17.67 mmol) was heated at 175°C for 10 hours. After cooling to room temperature, the reaction mixture was dissolved in methanol, and the solution was added dropwise to ether. The resulting product was filtered off and washed three times with ether, yielding a yellow solid. $^1$H NMR (400 MHz, Methanol-$d_4$) δ 8.35 (dd, $J$ = 8.2, 1.2 Hz, 1H), 8.31 (d, $J$ = 8.5 Hz, 1H), 7.96 (ddd, $J$ = 8.5, 7.2, 1.2 Hz, 1H), 7.85 (ddd, $J$ = 8.2, 7.2, 1.0 Hz, 1H), 4.87 (q, $J$ = 7.4 Hz, 2H), 3.28-3.24 (m, 3H), 1.63 (t, $J$ = 7.4 Hz, 3H). HRMS (ESI): $m/z$ calcd for $C_{10}H_{12}NS^+$: 178.05850 [$M$]$^+$; found: 178.05852.

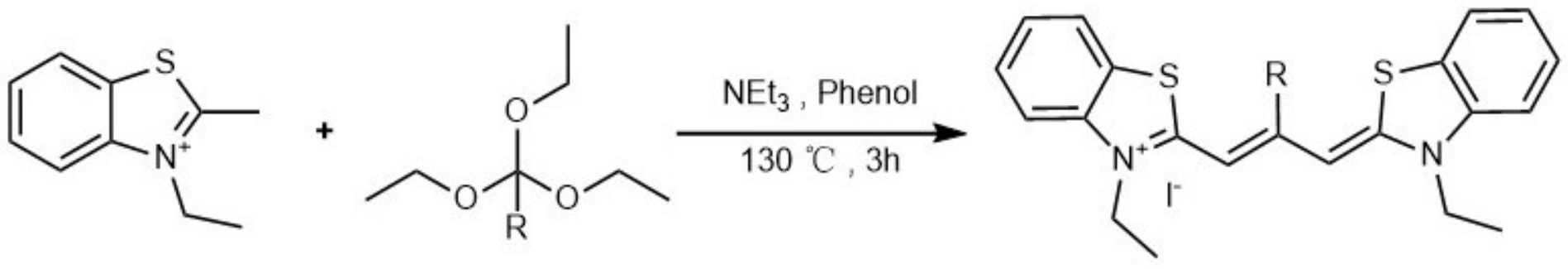

3 4

R= H BT-Cy-0
Me BT-Cy-1
Et BT-Cy-2

**Scheme S2** Synthetic scheme of **BT-Cy-0**, **BT-Cy-1**, and **BT-Cy-2**

**3-ethyl-2-((1E,3Z)-3-(3-ethylbenzo[d]thiazol-2(3H)-ylidene)prop-1-en-1-yl)benzo[d]thiazol-3-ium iodide (BT-Cy-0)**

A mixture of compound 2 (0.3054 g, 1.001 mmol), triethyl orthoformate (0.1582 g, 1.067 mmol), phenol (0.8721 g, 9.267 mmol), and triethylamine (0.2490 g, 2.481 mmol) was heated at 130°C in a thick-walled, pressure-resistant tube for 3 hours. After cooling to room temperature, the reaction mixture was dissolved in methanol and subjected to solvent exchange with ether in a chamber. The resulting product was filtered off and washed three times with ether. The product was then dissolved in methanol, and the solution was added dropwise to ether. The product that precipitated in the ether was filtered off and washed three times with ether to yield a dark blue solid. $^1$H NMR (400 MHz, Methanol-$d_4$) δ 7.99 (t, *J* = 12.8 Hz, 1H), 7.84 (d, *J* = 7.9 Hz, 2H), 7.65 (d, *J* = 8.3 Hz, 2H), 7.57 (t, *J* = 7.8 Hz, 2H), 7.41 (t, *J* = 7.6 Hz, 2H), 6.55 (d, *J* = 12.7 Hz, 2H), 4.41 (q, *J* = 7.2 Hz, 4H), 1.45 (t, *J* = 7.2 Hz, 6H). HRMS (ESI): *m/z* calcd for $C_{21}H_{21}N_2S_2^+$: 365.11407 [*M*]$^+$; found: 365.11447.

**3-ethyl-2-((1E,3Z)-3-(3-ethylbenzo[d]thiazol-2(3H)-ylidene)-2-methylprop-1-en-1-yl)benzo[d]thiazol-3-ium iodide (BT-Cy-1)**

A mixture of compound 2 (0.3800 g, 1.245 mmol), triethyl orthoacetate (0.2053 g, 1.266 mmol), phenol (0.5397 g, 5.735 mmol), and triethylamine (0.2417 g, 2.388 mmol) was heated at 130°C in a thick-walled, pressure-resistant tube for 3 hours. After cooling to room temperature, the reaction mixture was dissolved in methanol and subjected to solvent exchange with ether in a chamber. The precipitated product was filtered off and washed three times with ether. The product was then dissolved in methanol, and the solution was added dropwise to ether. The product that precipitated in the ether was filtered off and washed three times with ether, yielding a purple solid. $^1$H NMR (400 MHz, DMSO-d6) δ 8.09 (d, J = 7.9 Hz, 2H), 7.84 (d, J = 8.3 Hz, 2H), 7.63 (t, J = 7.8 Hz, 2H), 7.44 (t, J = 7.6 Hz, 2H), 6.62 (s, 2H), 4.52 (q, J = 7.0 Hz, 4H), 2.62 (s, 3H), 1.40 (t, J = 7.1 Hz, 6H). HRMS (ESI): m/z calcd for $C_{22}H_{23}N_2S_2^+$: 379.12972 [M]+; found: 379.12977.

**3-ethyl-2-((E)-2-((Z)-(3-ethylbenzo[d]thiazol-2(3H)-ylidene)methyl)but-1-en-1-yl)benzo[d]thiazol-3-ium iodide (BT-Cy-2)**

A mixture of compound 2 (0.5130 g, 1.681 mmol), triethyl orthopropionate (0.8277 g, 4.696 mmol), phenol (0.4427 g, 4.704 mmol), and triethylamine (0.4346 g, 4.295 mmol) was heated at 130°C in a thick-walled, pressure-resistant tube for 3 hours. After cooling to room temperature, the reaction mixture was dissolved in methanol and subjected to solvent exchange with ether in a chamber. The precipitated product was filtered off and washed three times with ether. The product was then dissolved in methanol, and the solution was added dropwise to ether. The product that precipitated in the ether was filtered off and washed three times with ether, yielding a dark purple solid. $^1$H NMR (600 MHz, Methanol-d4) δ 8.03 (dd, J = 71.4, 8.1 Hz, 2H), 7.80 (dd, J = 50.2, 8.3 Hz, 2H), 7.74 – 7.62 (m, 2H), 7.56 – 7.43 (m, 2H), 6.62 (d, J = 107.8 Hz, 2H), 4.64 (dq, J = 66.9, 7.2 Hz, 4H), 3.25 – 3.00 (m, 2H), 1.65 – 1.51 (m, 6H), 1.47 – 1.39 (m, 3H).HRMS (ESI): m/z calcd for $C_{23}H_{25}N_2S_2^+$: 393.14537 [M]+; found: 393.14498.

## 3. Structural Identification of the Dimer Formation of BT-Cy Probes

For each BT-Cy probe, temperature-dependent $^1$H-NMR experiments were performed to observe the behavior of the aromatic hydrogens and meta-substituent hydrogens as the temperature varied (Figure S1). At 213 K, each hydrogen showed only one set of corresponding signals, which were attributed to the dimer formed by the probe molecules. This is consistent with the fact that lower temperatures favor dimer formation.

As the temperature increased, especially from 253 K to 293 K, an additional set of signals gradually appeared for each hydrogen, with these new signals becoming sharper and more intense as the temperature rose. This indicates that, as the temperature increased, part of the dimer dissociated into monomers, confirming the temperature-dependent nature of the dimer-monomer equilibrium.

Additionally, 2D NOE NMR experiments were conducted on all three probes (Figure S2), and some typical inter-molecular NOE signals (red boxes) were observed in the NOE spectra of each probe. These signals further confirm the formation of dimers.

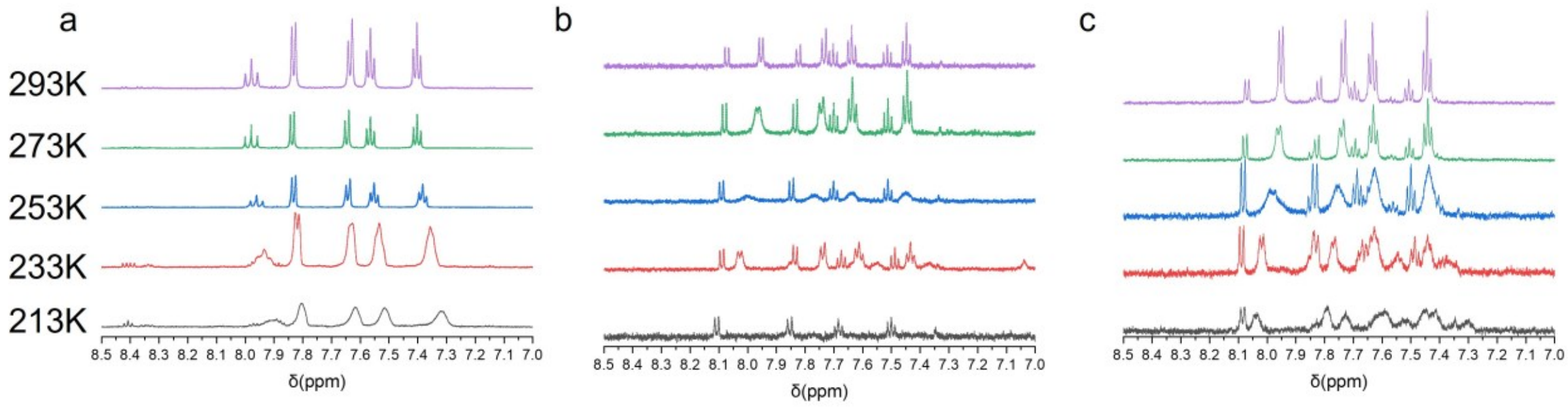

**Figure S1** Variable temperature $^1$H-NMR spectra from 213 K to 293 K of **a BT-Cy-0**, **b BT-Cy-1**, and **c BT-Cy-2**.

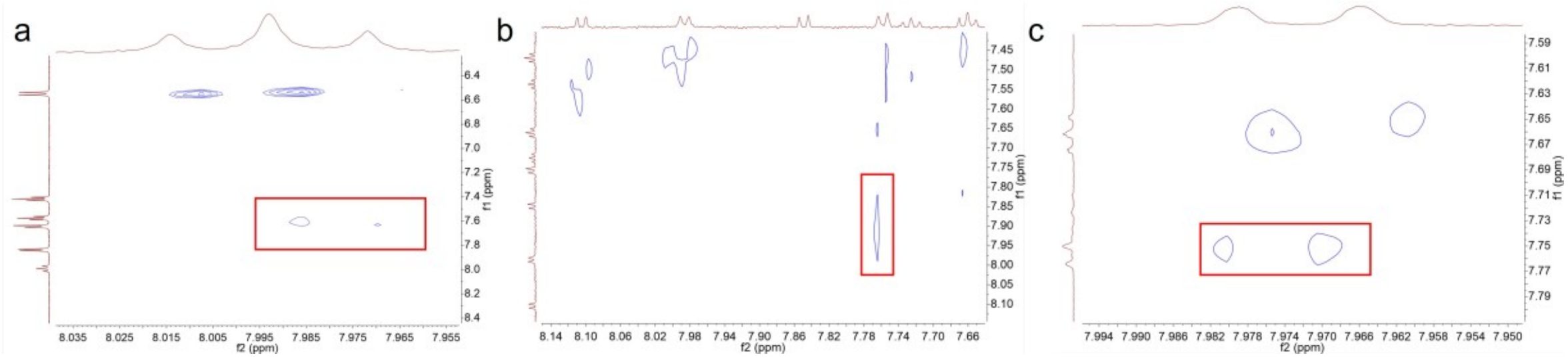

**Figure S2** NOESY spectra of **a BT-Cy-0**, **b BT-Cy-1**, and **c BT-Cy-2**.

## 4. Comparison of the Dimer Thicknesses of BT-Cy Probes by DOSY NMR

The dimer thickness of the BT-Cy probes was compared semi-quantitatively using DOSY NMR spectroscopy. As shown in Figure S3, the diffusion coefficients (D, y-axis) of each probe system, along with those of the solvent and TMS reference, exhibit notable differences, indicating that the probes form dimers under the experimental conditions.

To better compare the sizes of the different probes, we used the diffusion coefficient of TMS ($D_{TMS}$) as a reference and calculated the $D_{TMS}/D_{Probe}$ ratio (Table S2). This approach allowed us to assess the relative sizes of the probe dimers. When comparing **BT-Cy-0** and **BT-Cy-1**, the $D_{TMS}/D_{Probe}$ ratio of **BT-Cy-1** was higher, suggesting that its dimer is larger than that of **BT-Cy-0**. In contrast, the $D_{TMS}/D_{Probe}$ ratio of BT-**Cy-1** and **BT-Cy-2** were nearly identical, indicating that their dimer sizes cannot be resolved under the current experimental conditions.

The resolution of DOSY in terms of molecular size is approximately 0.5 Å.[3] According to our molecular docking calculations, **BT-Cy-1** is 0.8 Å thicker than **BT-Cy-0**, which is reflected in the small but noticeable difference in their $D_{TMS}/D_{Probe}$ ratios. On the other hand, since **BT-Cy-2** is only 0.4 Å thicker than **BT-Cy-1**, the DOSY results do not show a significant difference in their $D_{TMS}/D_{Probe}$ ratios, making it difficult to distinguish their size difference.

Overall, the DOSY NMR experiment validated our probe design strategy: as the size of the meta-substituent increases, the dimer thickness of the probes also increases.

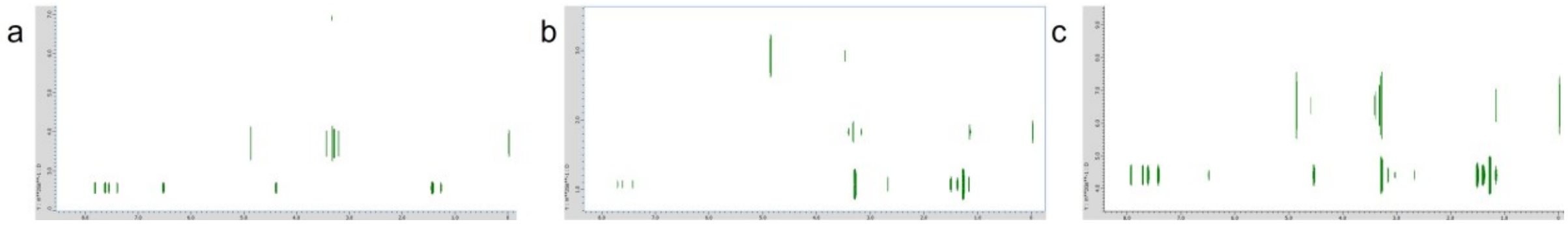

**Figure S3** DOSY NMR spectra of **a BT-Cy-0**, **b BT-Cy-1**, **c BT-Cy-2**.

**Table S2** Diffusion Coefficients and TMS/Probe Ratios for BT-Cy Probes

| Diffusion coefficient (× $10^{-9}$ $m^2/s$) | BT-Cy-0 | BT-Cy-1 | BT-Cy-2 |
|---|---|---|---|
| TMS | 3.709 | 1.837 | 6.896 |
| Probe | 2.579 | 1.076 | 4.115 |
| TMS/Probe ratio | 1.44 | 1.71 | 1.68 |

## 5. Confirming the Groove-Binding Mode of the Probes to G4 and iMs

### Probe Binding to G4

The binding mode of the probes to G4 structures has been confirmed through previous literature, where it was shown that the dimer of DMSB binds specifically to the groove of G4[4]. DMSB shares a highly similar structure to the probes used in this study, suggesting that our probes likely follow the same groove-binding pattern when interacting with G4 structures.

### Probe Binding to iM

In this study, we experimentally confirmed the groove-binding mode of the **BT-Cy-1** probe to the human telomeric iM structure by introducing bromination at each cytosine position along the groove.

Figure S4**a** shows the CD spectra of the brominated i-hTel sequences. It can be observed that the CD spectra of most of the brominated sequences remain almost unchanged, indicating that bromination does not affect the formation of the iM structure. However, for the 8-Br and 9-Br sequences, the positive peaks in the CD spectra shift towards shorter wavelengths, though they still lie within the characteristic signal region for iM structures. [5]

Fluorescence intensity measurements (ΔFI) showed that bromination at all positions resulted in a fluorescence decrease ranging from 40% to 60%, suggesting that steric hindrance in the groove generally affects the probe's ability to bind to the iM structure (Figure S4**b**).

Interestingly, the fluorescence changes at the 2-Br and 8-Br positions were relatively weaker compared to other positions (orange columns). This suggests that bromination at these two sites has a smaller impact on the probe-iM binding interaction. Both 2-Br and 8-Br are located between the C-C$^+$ base pairs within the iM structure, and we hypothesize that these positions correspond to regions with less steric hindrance, potentially aligning with the cyanoethyl chain of the probe, which allows for a more favorable binding interaction.

These findings further support the conclusion that the binding mode of **BT-Cy-1** to the iM groove is groove-binding, and the steric hindrance caused by bromination can modulate the binding efficiency depending on the position of modification.

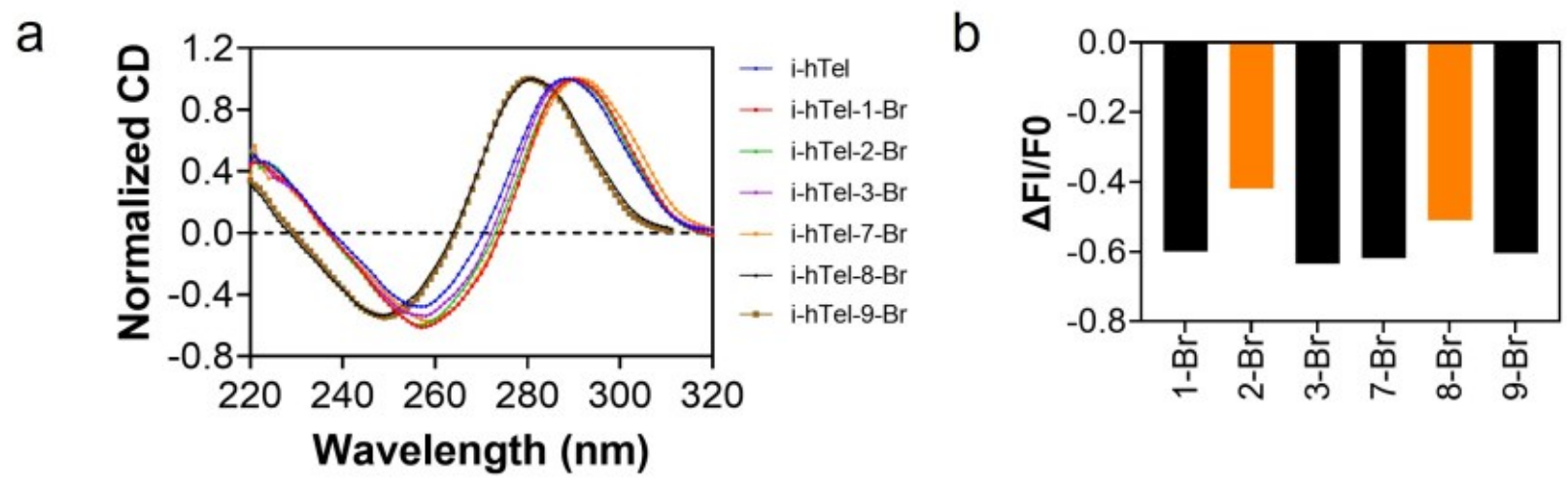

**Figure S4** Verification of groove binding. **a** Structural verification of wild-type and brominated modifications of i-hTel (5 μM) in CD spectra; **b** Percentage decrease in fluorescence intensity of the brominated sequence compared to the wild-type sequence upon binding of **BT-Cy-1** (5 μM).

## 6. UV Spectra of BT-Cy Probes Binding dsDNA

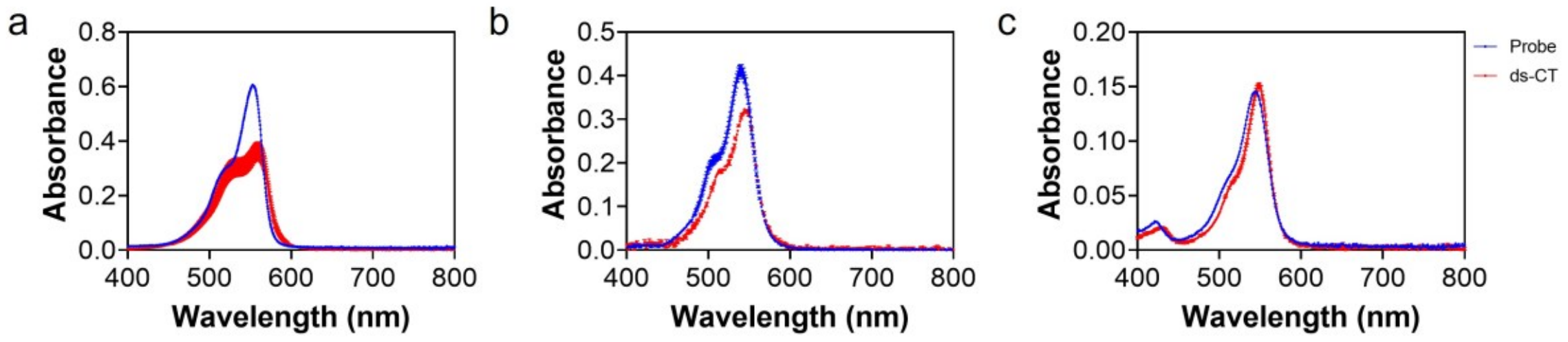

**Figure S5** UV spectra of dsDNA (36 μg/mL) interaction with three probes (5 μM):

**a BT-Cy-0**, **b BT-Cy-1**, **c BT-Cy-2**.

## 7. Verification of DNA Structures

As is shown at Figure S6, the positive absorption at 290 nm and negative absorption around 260 nm are characteristic features of classical iM structures.[5] The positive absorption peak at 260 nm and the negative absorption peak at 240 nm are characteristics of parallel G4,[6] while the positive absorption peak at 290 nm, the shoulder peak at around 270 nm, and the negative peak at 240 nm are characteristics of hybrid G4.[7]

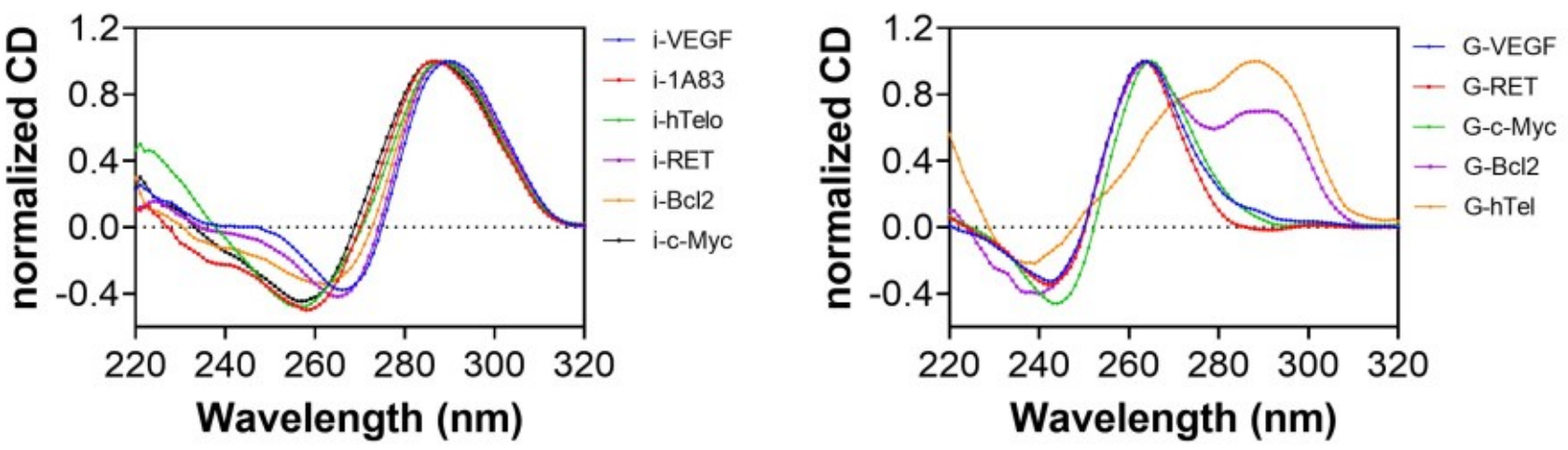

**Figure S6** CD spectra of DNA (5 μM): iM (left) and G4 (right).

## 8. Raw Fluorescence Spectra of BT-Cy Probes Binding to Different DNAs.

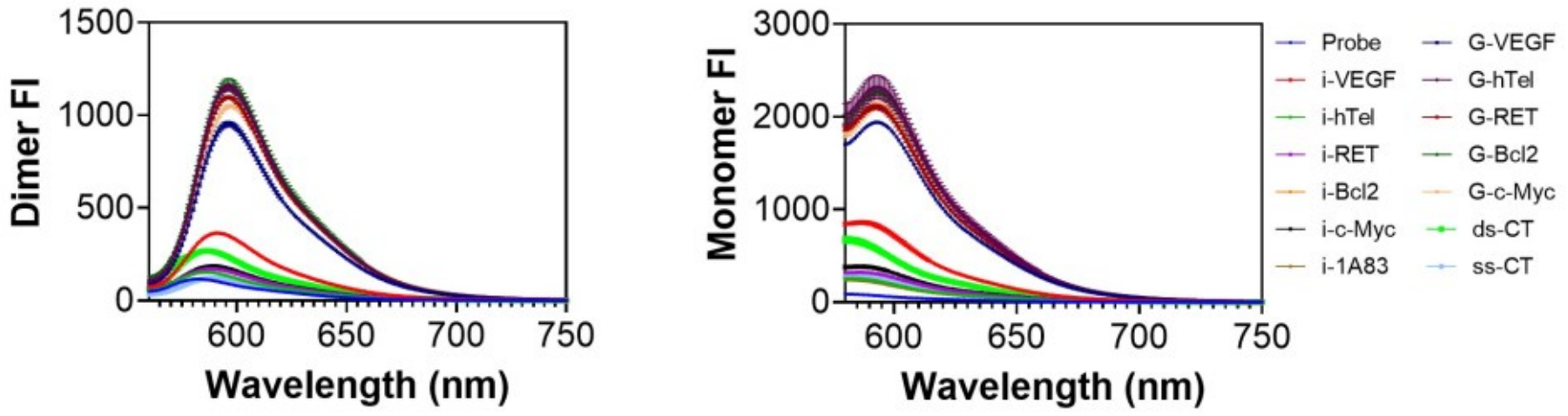

**Figure S7** Fluorescence spectra of **BT-CY-0** (5 μM) dimer (left) and monomer (right) binding DNAs (5 μM).

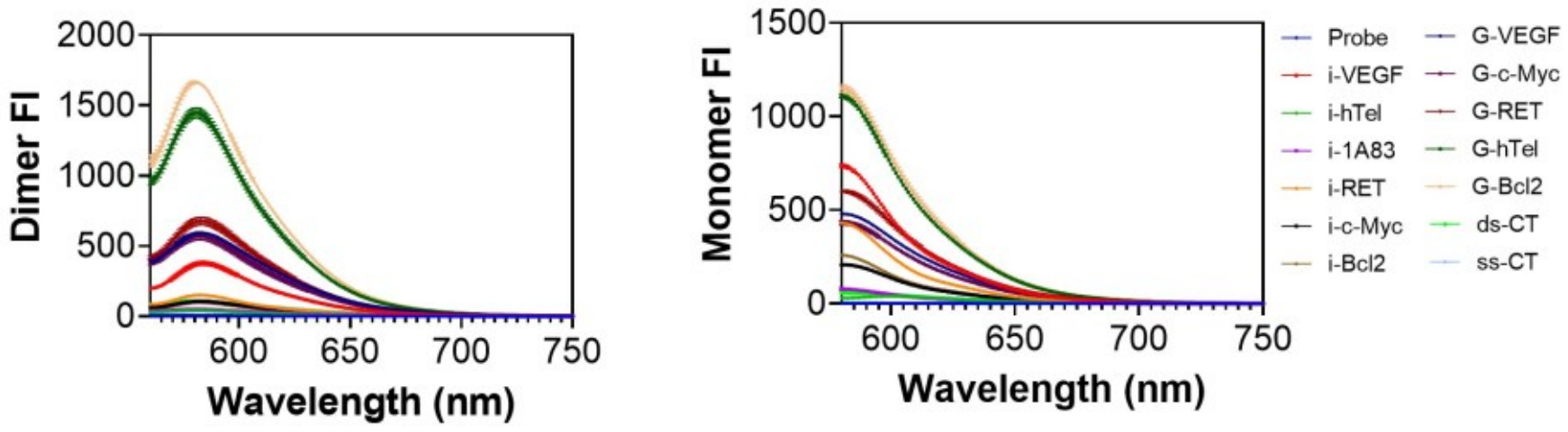

**Figure S8** Fluorescence spectra of **BT-CY-1** (5 μM) dimer (left) and monomer (right) binding DNAs (5 μM).

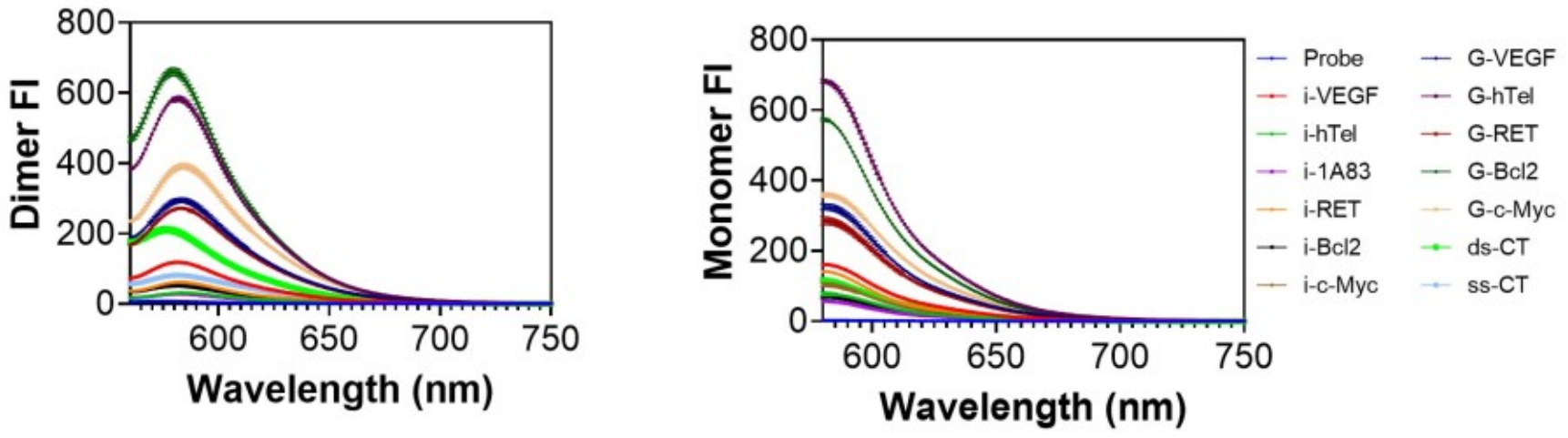

**Figure S9** Fluorescence spectra of **BT-CY-2** (5 μM) dimer (left) and monomer (right) binding DNAs (5 μM).

## 9. G4 and iM Structural Transitions and BT-Cy-1 Fluorescence Spectra

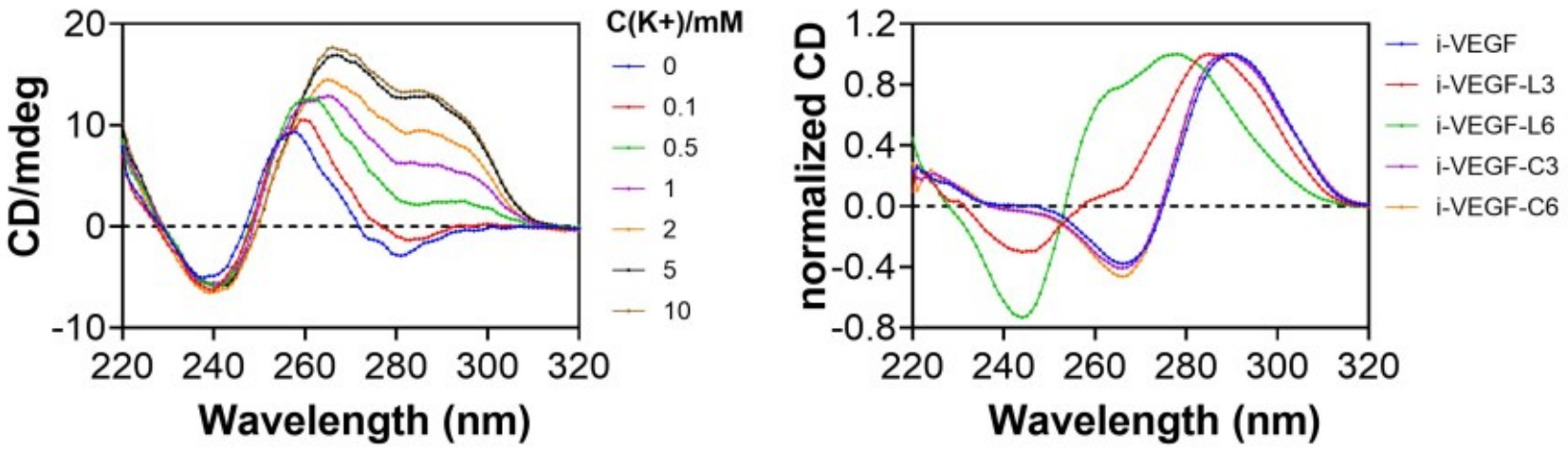

**Figure S10** CD spectra of G-hTel (5 μM) at different potassium ion concentrations (0-10 mM) (left) and i-VEGF sequences with loop or groove elongation (5 μM) (right).

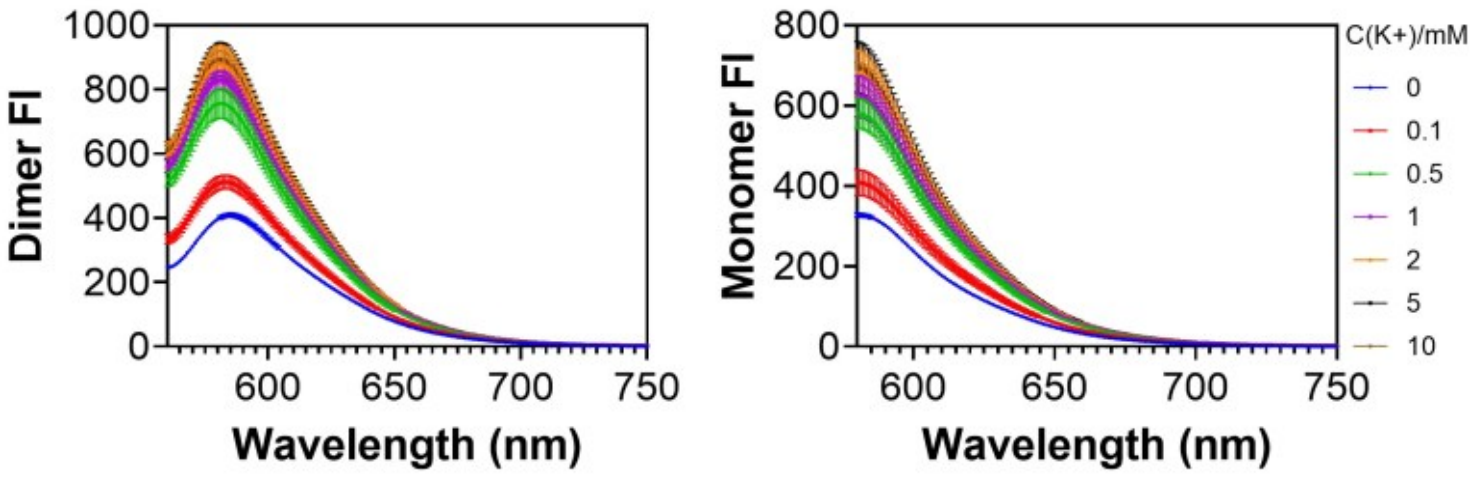

**Figure S11** Fluorescence spectra of dimer (left) and monomer (right) of **BT-Cy-1** (5 μM) binding to G-hTel (5 μM) in Tris-HCl buffer with different potassium ion concentrations (0-10 mM).

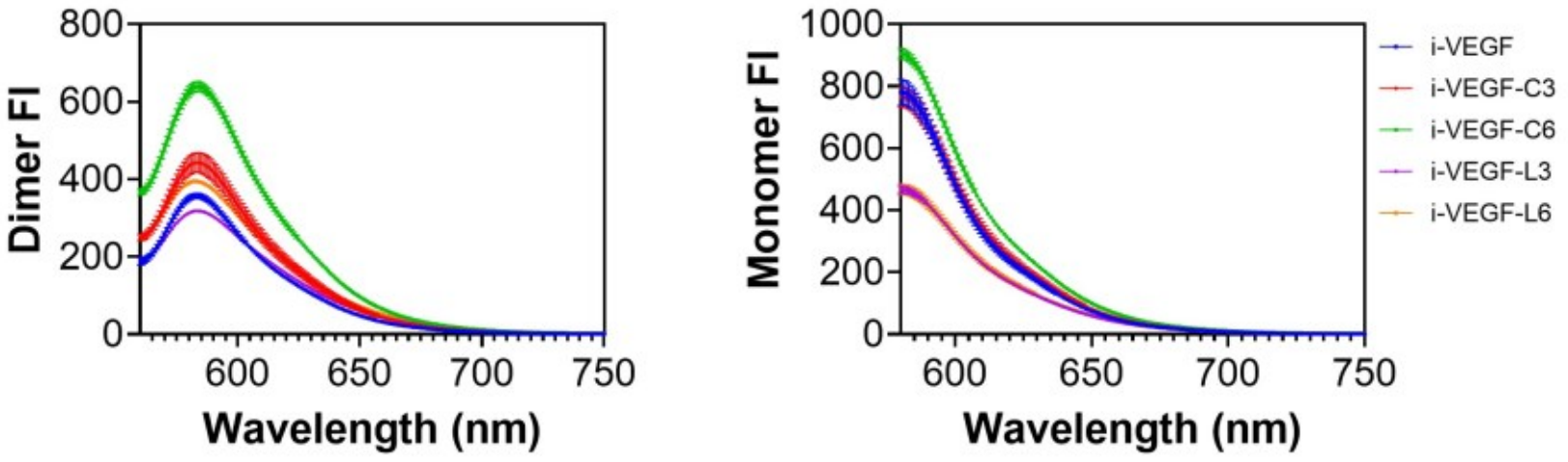

**Figure S12** Fluorescence spectra of **BT-Cy-1** (5 μM) dimer (left) and monomer (right) binding to wild-type i-VEGF and loop or groove elongation sequences (5 μM).

## 10. Detection Limits of BT-Cy-1 for G4 and iM under Interference from dsDNA

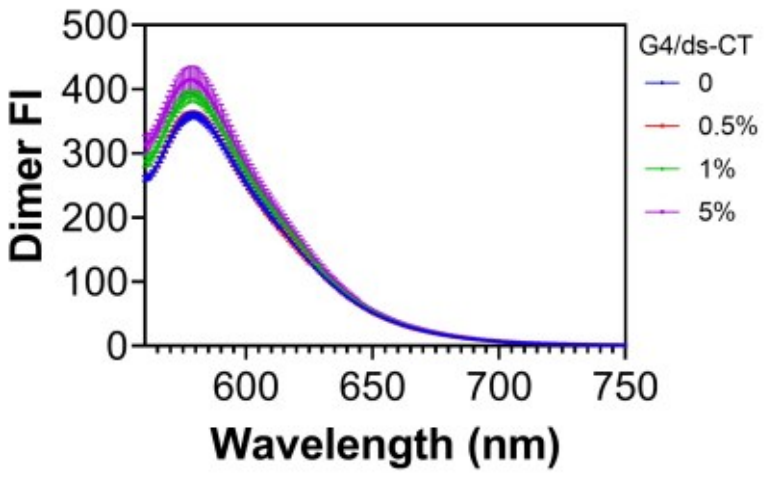

**Figure S13** Fluorescence spectra of **BT-Cy-1** (5 μM) dimer binding to G-VEGF at increasing proportions of G4 in dsDNA (36 μg/mL) (0%, 0.5%, 1%, and 5%).

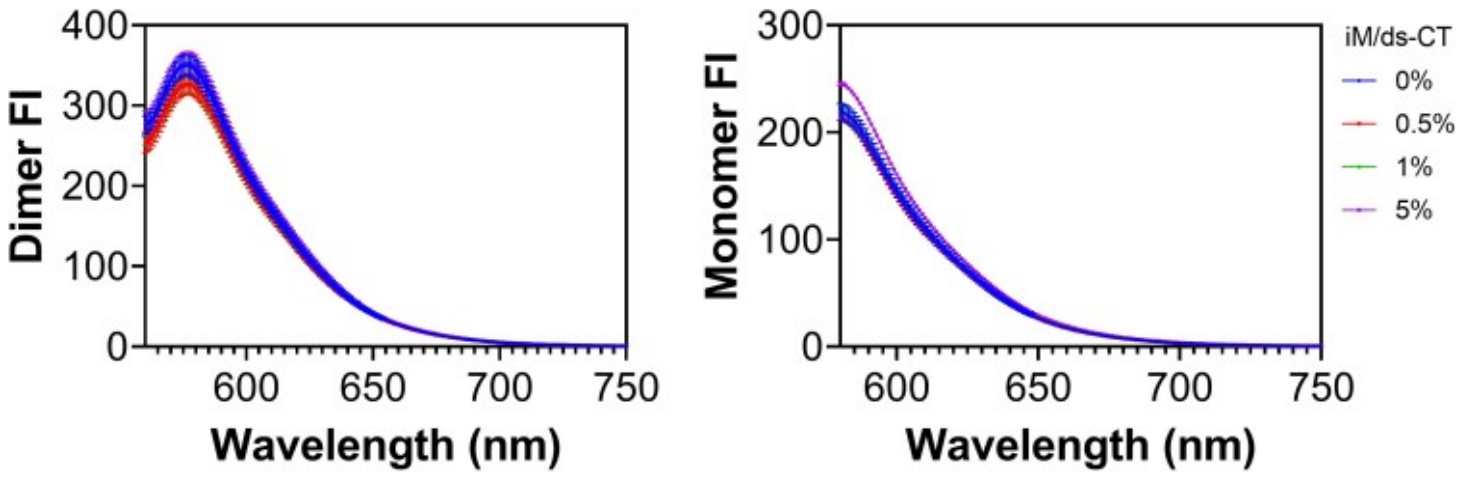

**Figure S14** Fluorescence spectra of **BT-Cy-1** (5 μM) dimer (left) and monomer (right) binding to i-VEGF at increasing concentrations of dsDNA (36 μg/mL) (0%, 0.5%, 1%, and 5%).

## 11. BT-Cy-1 Detection of G4 and iM Structures in Clinical Blood RNA Samples

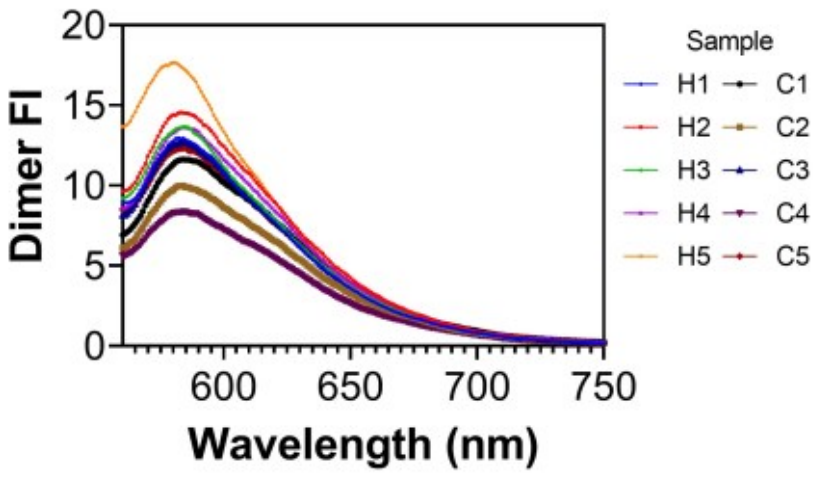

**Figure S15** Fluorescence spectra of **BT-Cy-1** (5 μM) dimer for detection of G4 levels in RNA (9 μg/mL) from clinical blood samples. H1-H5 are samples from healthy individuals and C1-C5 are samples from liver cancer patients.

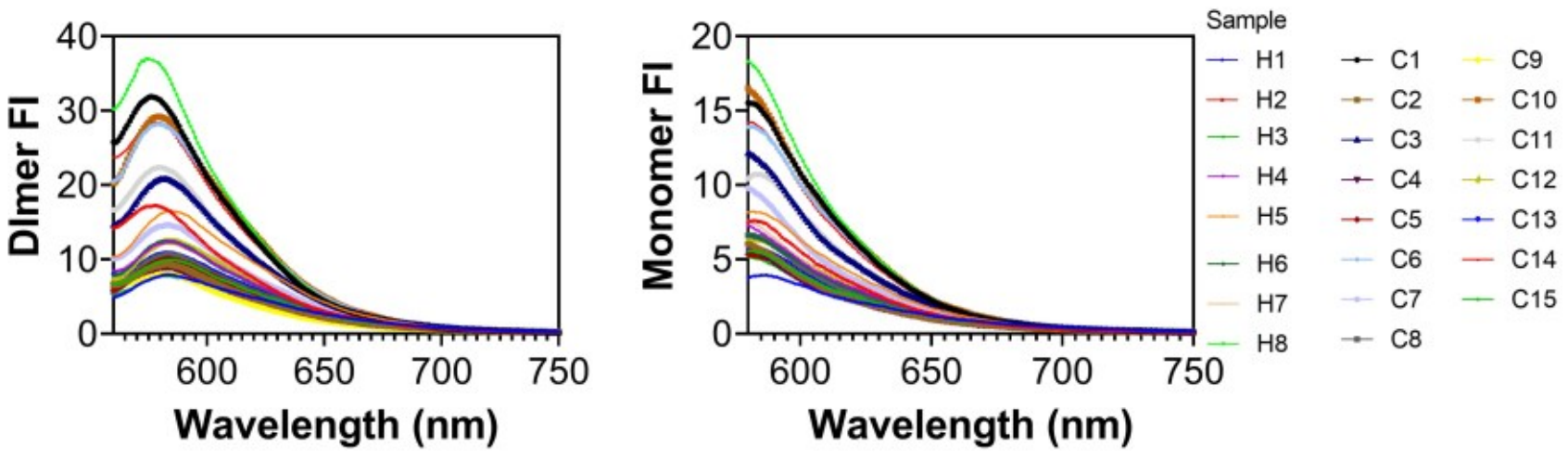

**Figure S16** Fluorescence spectra of **BT-Cy-1** (5 μM) dimer (left) and monomer (right) for detection of iM levels in RNA (9 μg/mL) from clinical blood samples. H1-H8 are samples from healthy individuals, and C1-C15 are samples from liver cancer patients.

## 12. NMR and MS Data

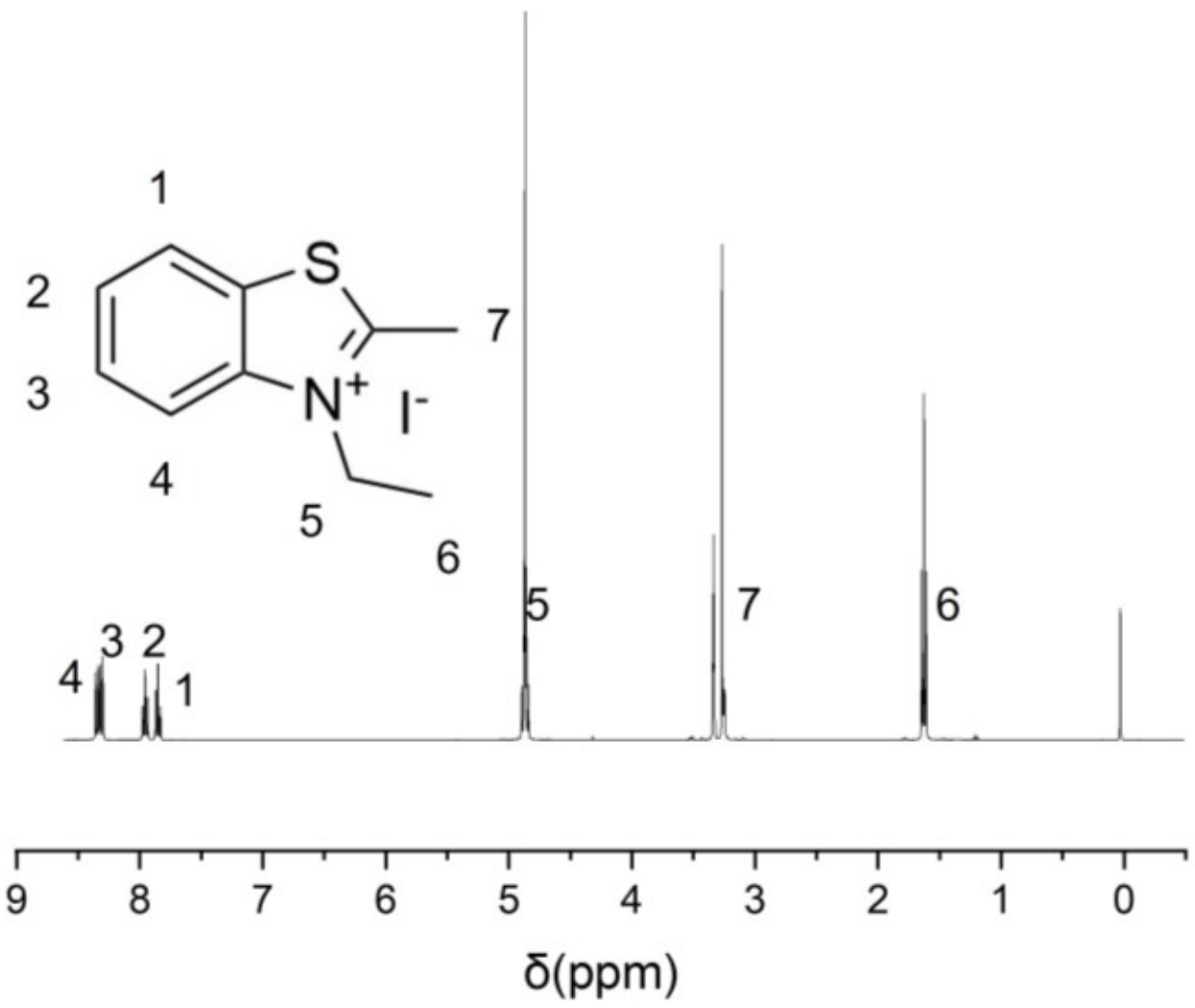

**Figure S17** $^1$H-NMR spectrum of compound 3 in $CD_3OD$ (400 MHz).

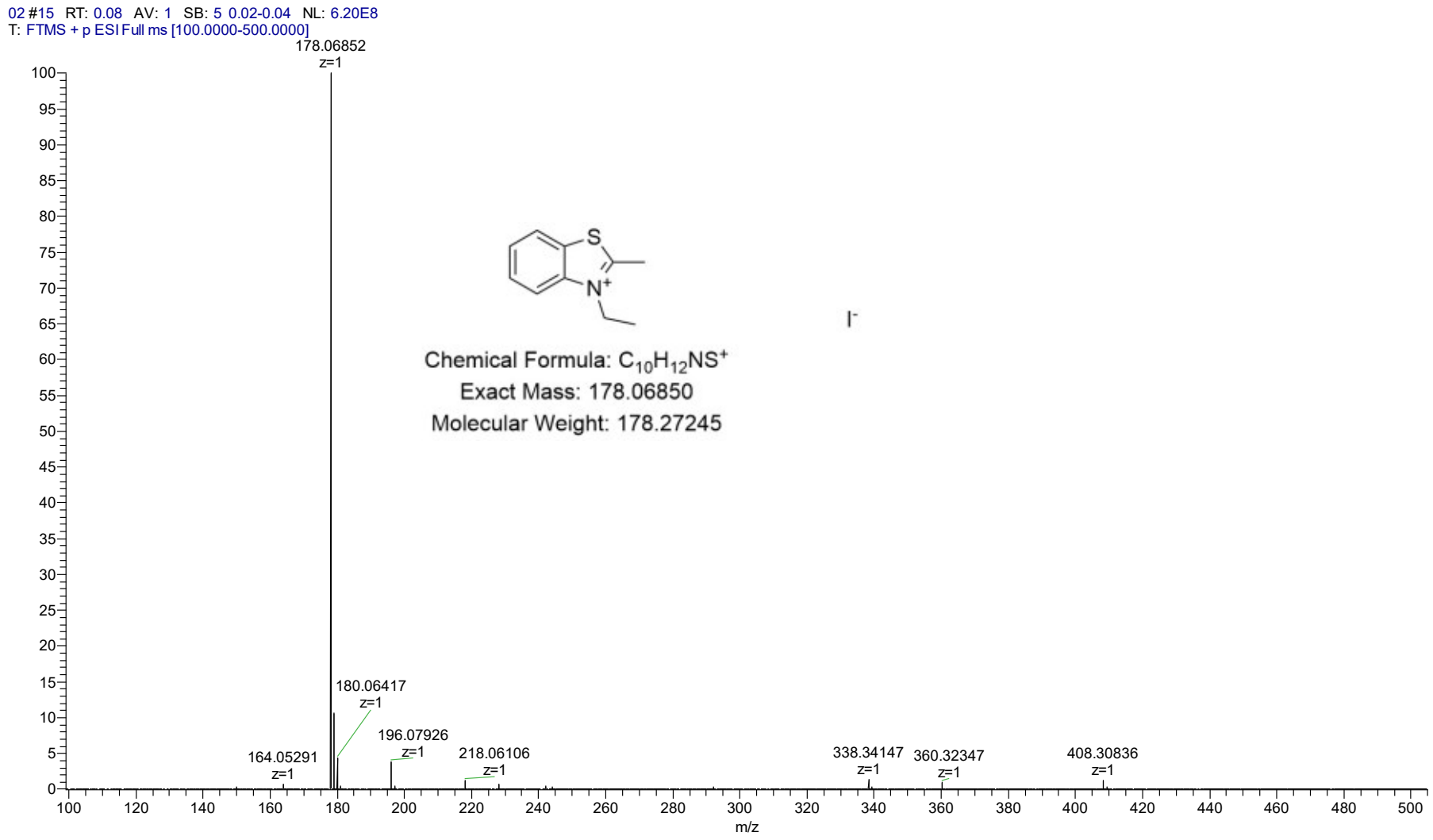

**Figure S18** HR-MS (H-ESI) spectrum of compound 3

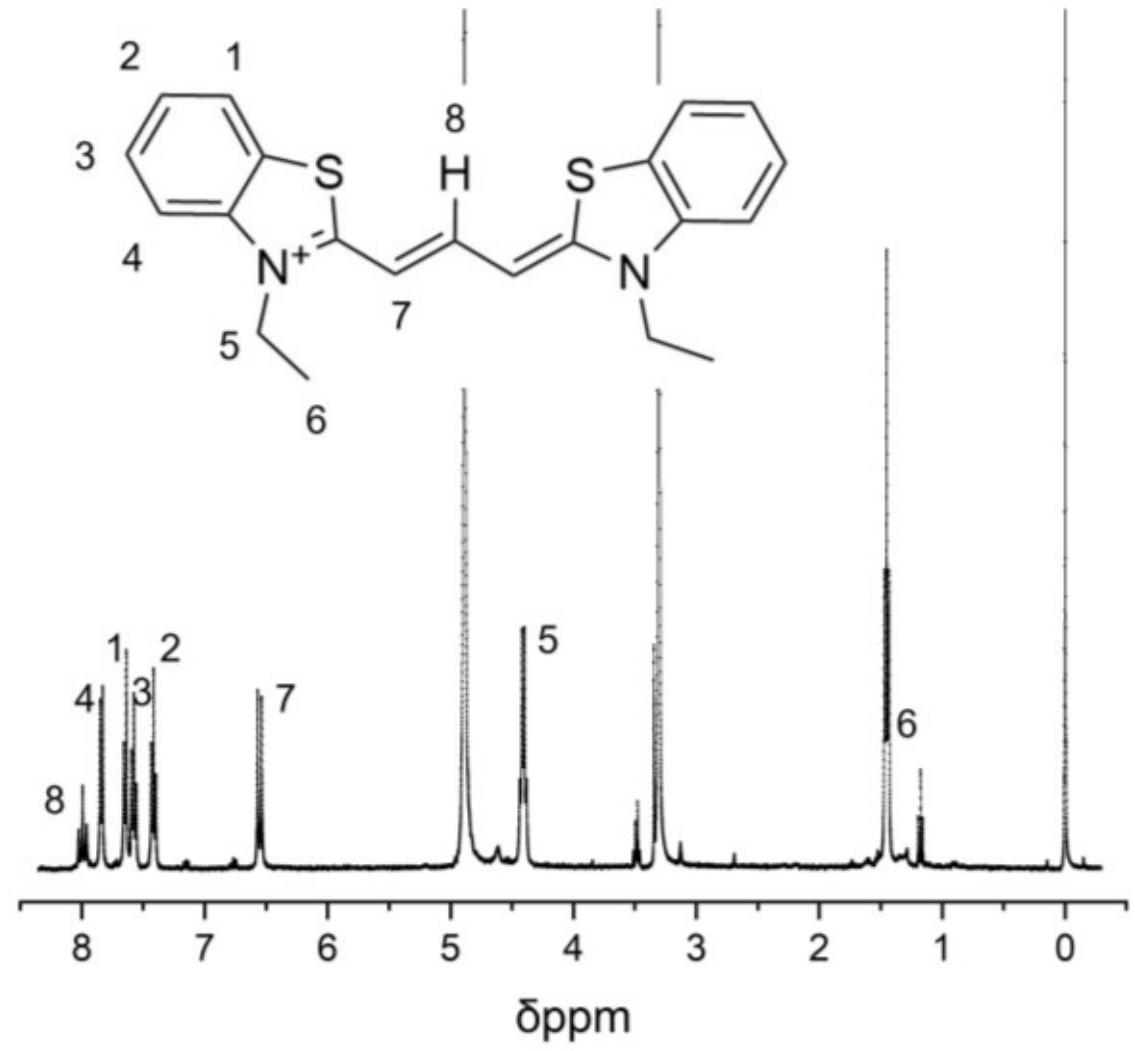

**Figure S19** $^{1}$H-NMR spectrum of **BT-Cy-0** in $CD_3OD$ (400 MHz).

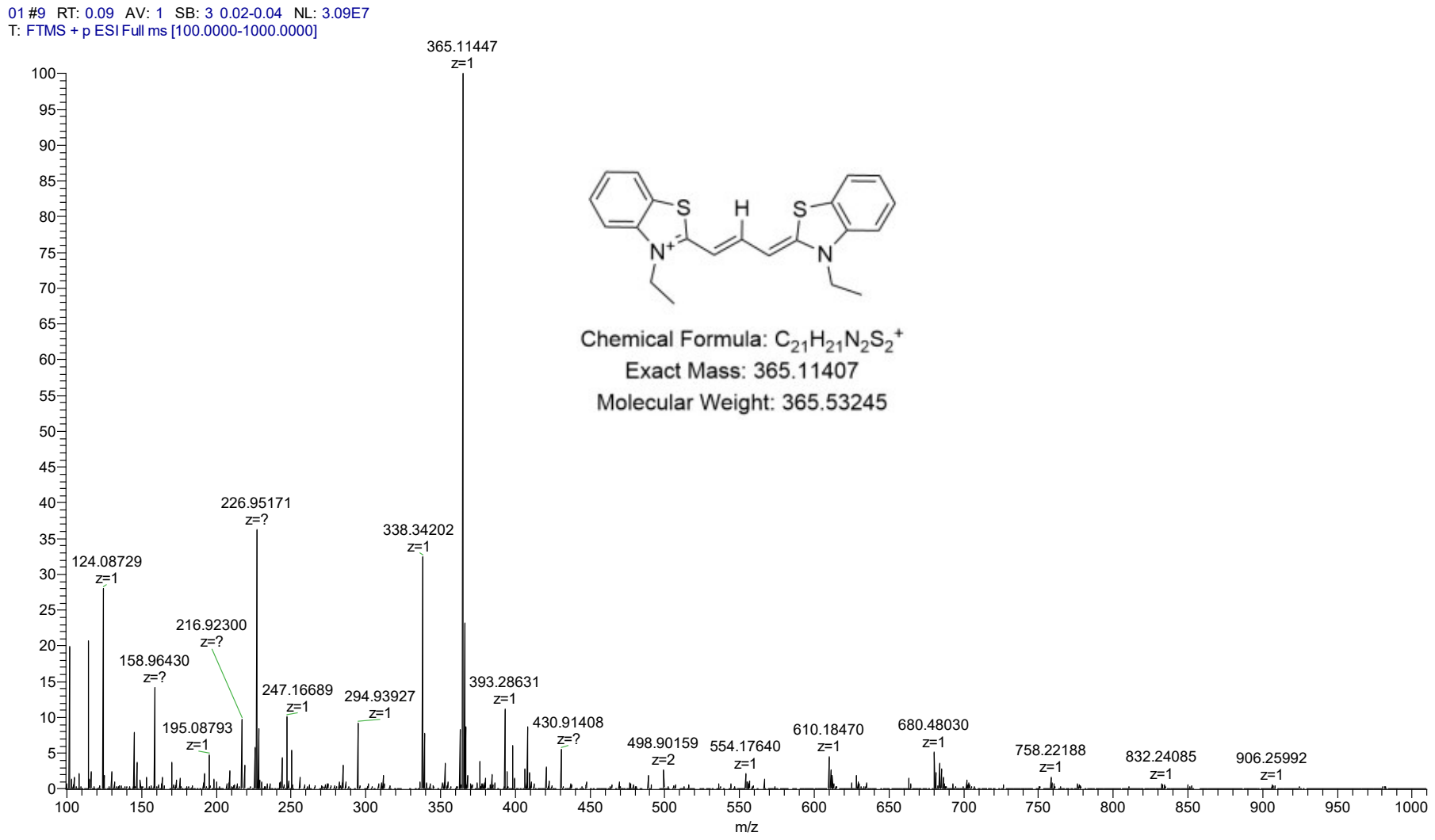

**Figure S20** HR-MS (H-ESI) spectrum of **BT-Cy-0**.

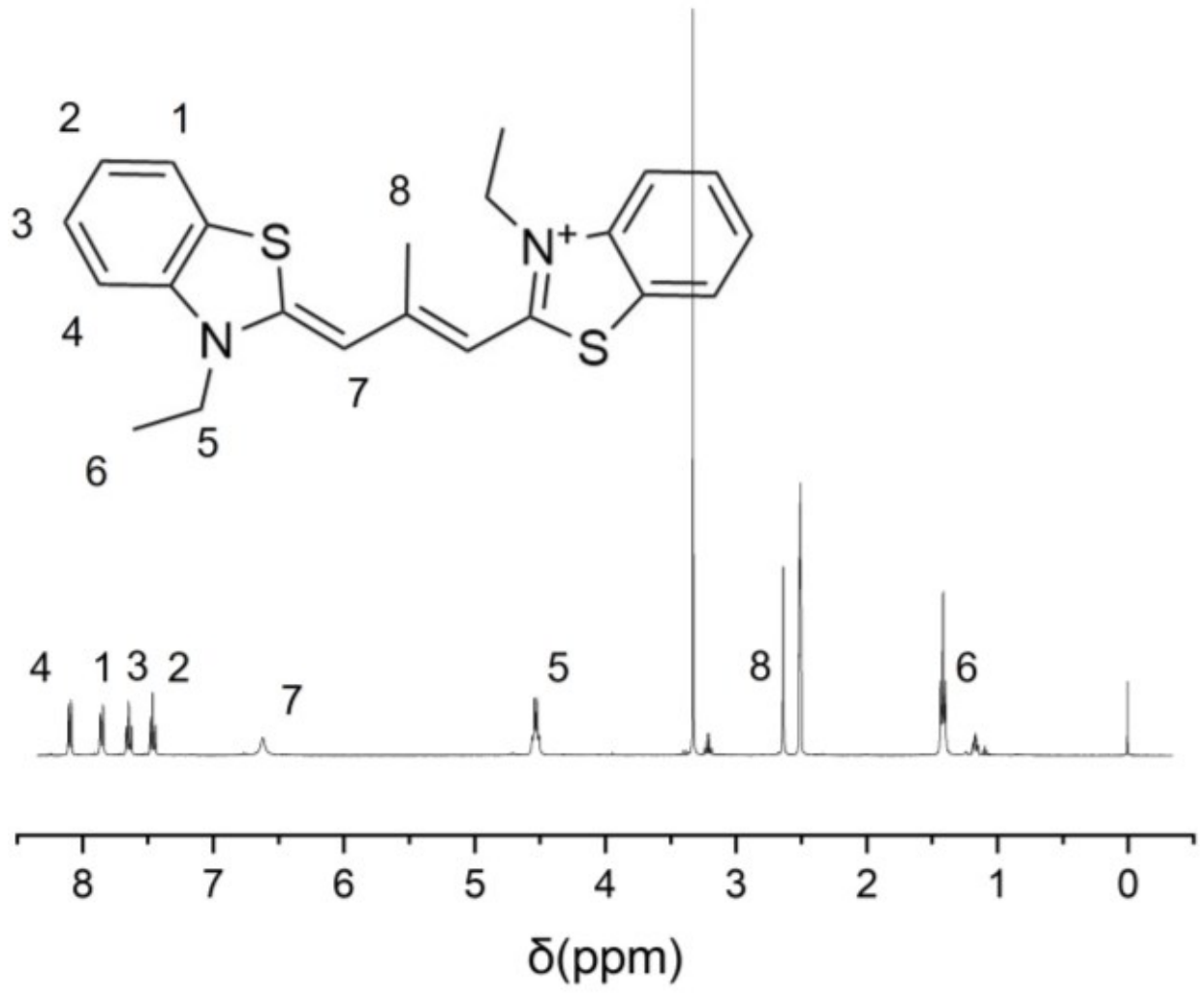

**Figure S21** [1]H-NMR spectrum of **BT-Cy-1** in $CD_3SOCD_3$ (400 MHz).

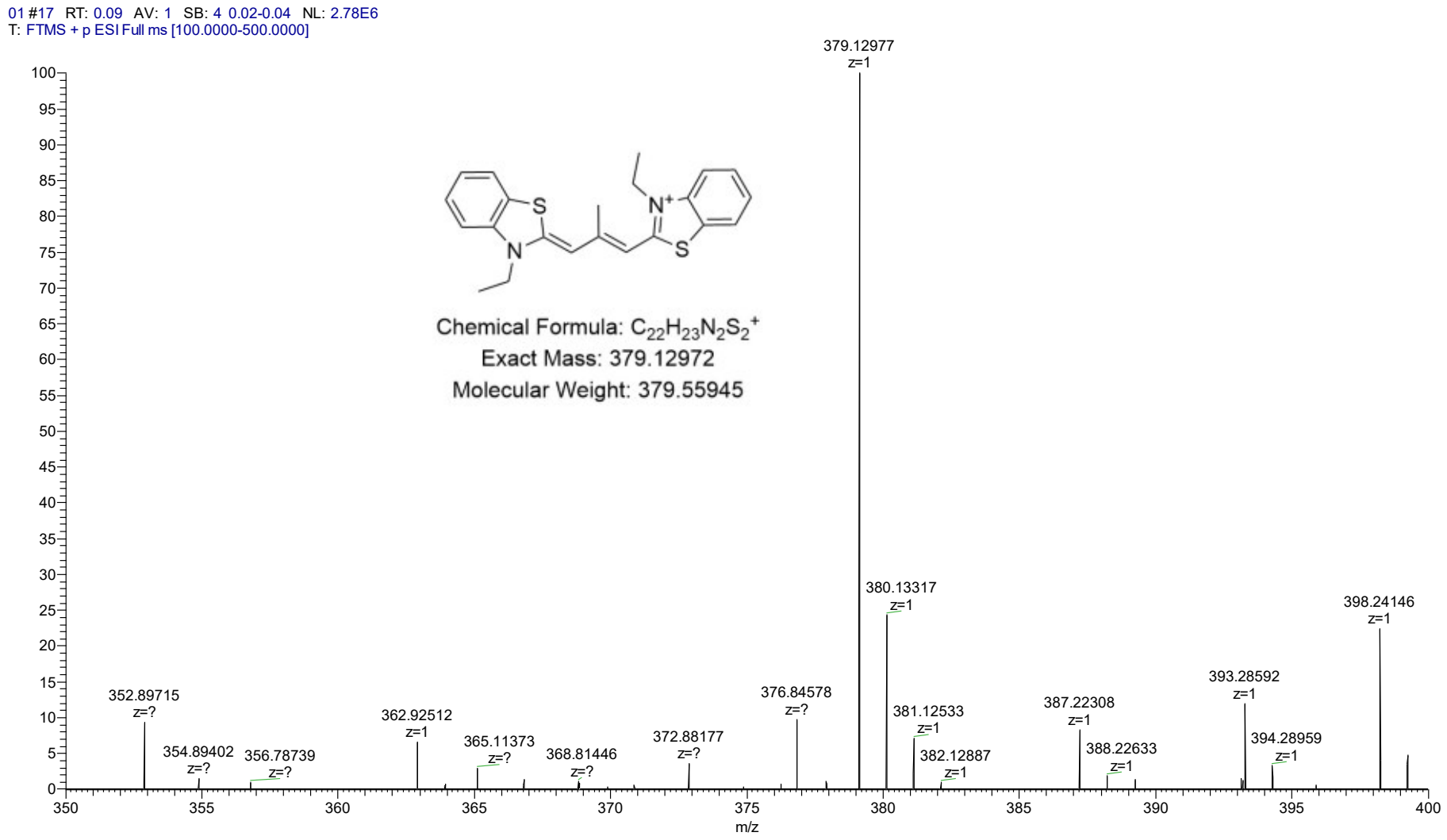

**Figure S22** HR-MS (H-ESI) spectrum of **BT-Cy-1**.

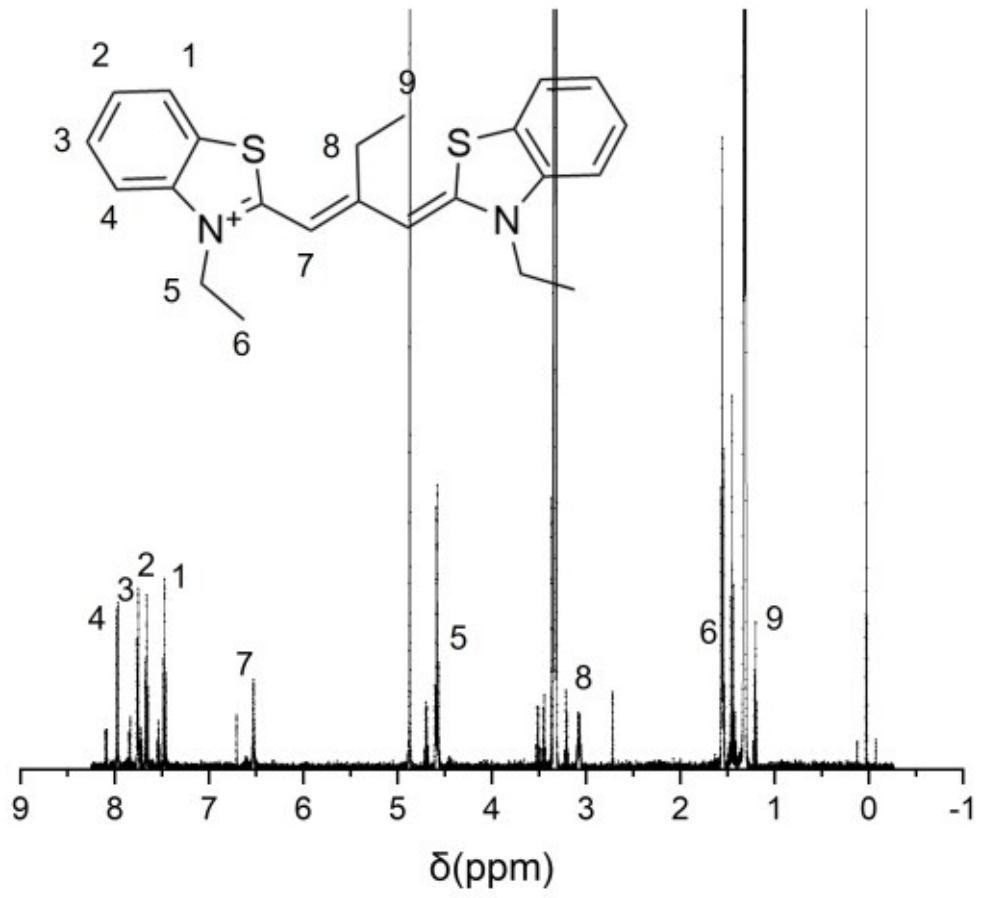

**Figure S23** $^{1}$H-NMR spectrum of **BT-Cy-2** in $CD_3OD$ (600 MHz).

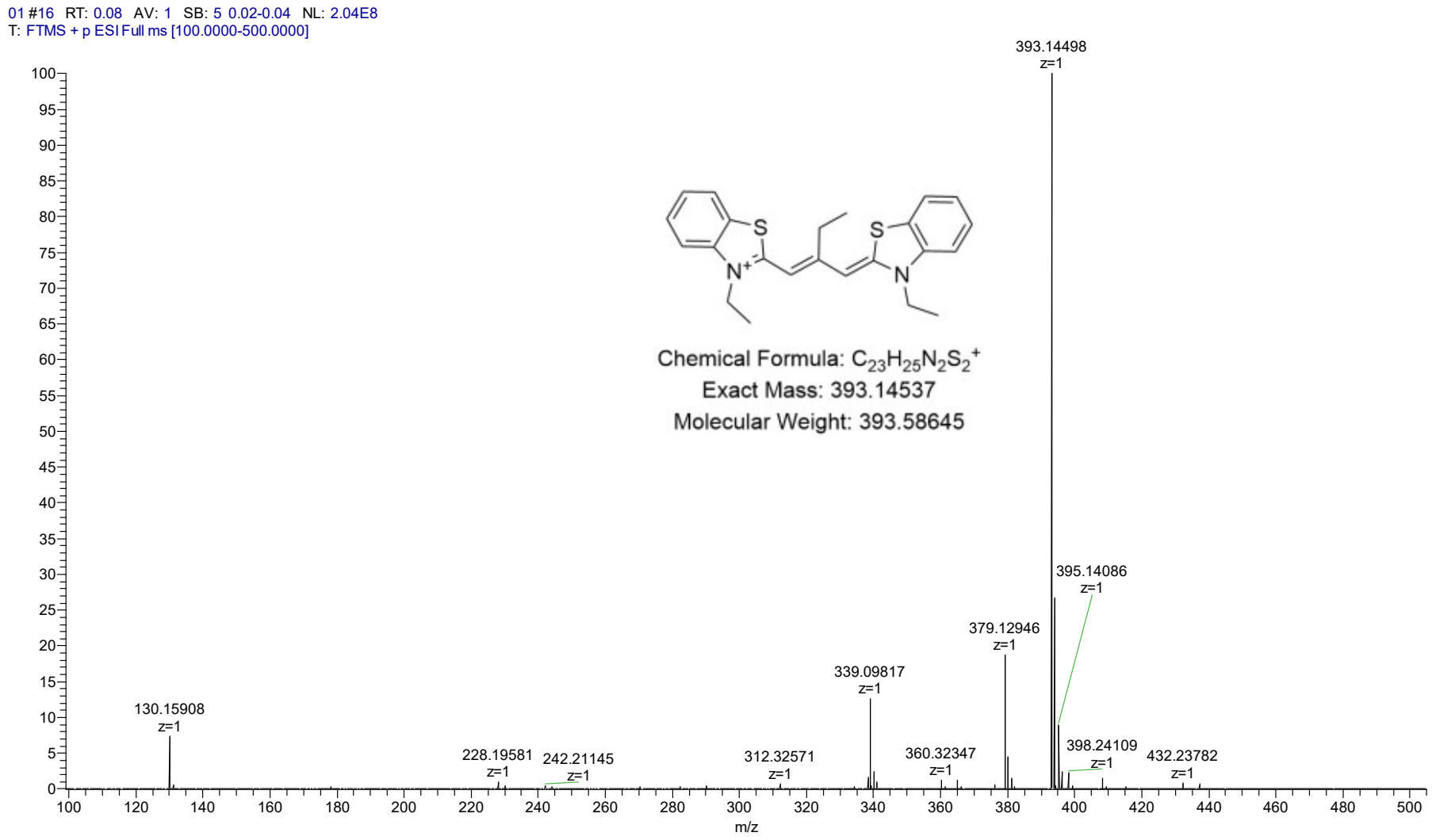

**Figure S24** HR-MS (H-ESI) spectrum of **BT-Cy-2**.